\documentclass[prd,twocolumn,floats,floatfix,nofootinbib]{revtex4}
\usepackage{graphicx}
\usepackage{dcolumn}
\usepackage{amssymb}
\usepackage{bm}

\def\spose#1{\hbox to 0pt{#1\hss}}

\def\lta{\mathrel{\spose{\lower 3pt\hbox{$\mathchar"218$}}
     \raise 2.0pt\hbox{$\mathchar"13C$}}}
\def\gta{\mathrel{\spose{\lower 3pt\hbox{$\mathchar"218$}}
     \raise 2.0pt\hbox{$\mathchar"13E$}}}
\newcommand{\be}{\begin{equation}}
\newcommand{\en}{\end{equation}}
\newcommand{\bea}{\begin{eqnarray}}
\newcommand{\ena}{\end{eqnarray}}

\newcommand{\dd}{\mbox{d}}

\newcommand{\eg}{\textsl{e.g.~}}
\newcommand{\etal}{\textsl{et al.~}}

\begin{document}
\title{Scalar and Vector Perturbations in Quantum Cosmological Backgrounds}

\author{Emanuel J. C. Pinho}
\email{emanuel@cbpf.br} \affiliation{ICRA - Centro Brasileiro de
Pesquisas F\'{\i}sicas -- CBPF, \\ rua Xavier Sigaud, 150, Urca,
CEP22290-180, Rio de Janeiro, Brazil}

\author{Nelson Pinto-Neto}
\email{nelsonpn@cbpf.br} \affiliation{ICRA - Centro Brasileiro de
Pesquisas F\'{\i}sicas -- CBPF, \\ rua Xavier Sigaud, 150, Urca,
CEP22290-180, Rio de Janeiro, Brazil}

\date{\today}

\begin{abstract}
Generalizing a previous work concerning cosmological linear tensor
perturbations, we show that the lagrangians and hamiltonians of
cosmological linear scalar and vector perturbations can be put in
simple form through the implementation of canonical transformations
and redefinitions of the lapse function, without ever using the
background classical equations of motion. In particular, if the
matter content of the Universe is a perfect fluid, the hamiltonian
of scalar perturbations can be reduced, as usual, to a hamiltonian
of a scalar field with variable mass depending on background
functions, independently of the fact that these functions satisfy
the background Einstein's classical equations. These simple
lagrangians and hamiltonians can then be used in situations where
the background metric is also quantized, hence providing a
substantial simplification over the direct approach originally
developed by Halliwell and Hawking.
\\


\end{abstract}

\maketitle

\section{Introduction}

In the theory of linear cosmological perturbations, simple evolution
equations for the perturbations have been obtained~\cite{MFB}.
Lagrangians and hamiltonians describing the dynamics of scalar,
vector, and tensor perturbations coming from the Einstein-Hilbert
lagrangian have been greatly simplified in different cosmological
scenarios under the assumption that the background metric satisfies
Einstein classical field equations, and after taking out space and
time total derivatives~\cite{MFB}. Once these simple lagrangians and
hamiltonians are obtained, the quantization of linear cosmological
perturbations becomes easy, with a quite simple interpretation: they
can be seen as quantum fields which behave essentially as scalar
fields with a time dependent effective mass. The time varying
background scale factor which is responsible for this ``mass'' acts
as a pump field~\cite{grish}, creating or destroying modes of the
perturbations. In this framework, one can assume an initial vacuum
state for the perturbations, yielding primordial perturbation
spectra which can be compared with observations. In the cosmological
inflationary scenario~\cite{inflation}, the resulting spectrum for
scalar perturbations is in good agreement with the
data~\cite{WMAP3}.

However, this state of affairs is rather incomplete: the
overwhelming majority of classical backgrounds possess an initial
singularity at which the classical theory is expected to break down,
and one needs to justify the initial conditions for inflation and
quantum perturbations. Hence, a full quantum treatment including the
background must be constructed. The first approach in this direction
was made in Ref.~\cite{halli}, where the canonical quantization of
the perturbations and background was implemented through the
derivation of the superhamiltonian constraint of the whole system
and its consequent Wheeler-DeWitt equation
$H(\hat{A},\hat{P_A},\hat{X},\hat{P_X})\Psi=0$, where $A$ and $P_A$
represent the phase space background variables, and $X$ and $P_X$
the perturbation phase space variables. They claim that the no
boundary proposal can set the initial conditions for inflation and
the vacuum initial state for the perturbations. Then, through the
imposition of the ansatz on the wave functional
$\Psi(A,X,t)=\varphi(A,t)\psi(A,X,t)$, they could manage to separate
the quantum effects in the background from the quantum
perturbations, where the wave function for the background
$\varphi(A,t)$ obeys an independent quantum minisuperspace
description where back reactions terms from the quantum
perturbations are negligible. The singularity is bypassed through an
euclidianization of space-time near it, and a consequent beginning
of time when (or where) the geometry passes from the euclidian
signature to the lorentzian one. The quantum perturbations are
described in the oscillatory part of the background wave function,
where a WKB approximation can be used. Then, the evolution of the
scale factor in time may be obtained through the equation
$\dot{a}\propto\partial S/\partial a$, where $S$ is a solution of
the classical Hamilton-Jacobi equation. Hence, this evolution is the
classical one, and we are back to a semiclassical description of the
perturbations.

In parallel to that, the possibility that the singularity could be
avoided through a bounce connecting the present expanding phase with
a preceding contracting phase has been explored. In this case, the
Universe is eternal: there is no beginning of time, nor horizons.
Many frameworks where bounces may occur have been
proposed~\cite{bounce,pbb,ekp}. These new features of the background
introduce a new picture for the evolution of cosmological
perturbations: vacuum initial conditions may now be imposed when the
Universe was very big and almost flat, and effects due to the
contracting and bouncing phases, which are not present in models
with a beginning of time, may change the subsequent evolution of
perturbations in the expanding phase. Because of that, the evolution
of cosmological perturbations in bouncing models has been cause of
intense debate~\cite{pertbounce}.

In the framework of quantum cosmology in minisuperspace models,
bouncing models had also been proposed where the bounce occurs due
to quantum effects in the
background~\cite{russo,tipler,lemos,pinto,pinto2,fabris,ash}. Some
approaches have used an ontological interpretation of quantum
mechanics, the Bohm-de Broglie~\cite{dbb} one, to interpret the
results~\cite{pinto,pinto2,fabris}). In this interpretation,
quantum Bohmian trajectories, the quantum evolution of the scale
factor $a_q(t)$ at zeroth order, can be defined through the
relation $\dot{a}\propto\partial S/\partial a$, where $S$ is now
the phase of the background wave function $\varphi(A,t)$ without
any approximation: it is not a solution of the classical
Hamilton-Jacobi equation. In fact it satisfies a modified
Hamilton-Jacobi equation derived from the Wheeler-DeWitt equation
for $\varphi(A,t)$, and hence $a_q(t)$ is not the classical
trajectory: in the regions where the quantum effects cannot be
neglected, the quantum trajectory $a_q(t)$ performs a bounce which
connect two asymptotic classical regions where the quantum effects
are negligible. One than has in hands a definite function of time
for the background, even at the quantum level, which realizes a
soft transition from the contracting phase to the expanding one.
Due to the results of Ref.~\cite{halli}, where the background
minisuperspace Wheeler-DeWitt equation for $\varphi(A,t)$ continue
to hold when quantum perturbations are present because back
reaction terms are negligible (which can also be justified through
other ansatz for the wave function or verified `a posteriori'),
this background quantum function $a_q(t)$ is sufficient to
describe the whole quantum features of the background. The natural
question to ask is what happens with the perturbations when it
passes through this well defined and regular quantum bounce. One
could then use the hamiltonian $H$ of Ref.~\cite{halli} to
investigate the evolution of quantum perturbations in this quantum
background. However, the structure of $H$ is rather complicated,
turning it difficult to obtain any detailed result about the
spectra of perturbations, specially the scalar ones. Also, a
simplification of $H$ using the zeroth order classical equations,
as done in Ref.~\cite{MFB} and described in the beginning of this
section, is not possible because the background is also quantized
and it does not satisfy the classical Einstein's equations. This
state of affairs motivated us to find a way to simplify the
hamiltonian of Ref.~\cite{halli}, without ever recurring to the
background classical equations, and apply it to these quantum
systems.

Recently, we have managed to put the hamiltonian of tensor
perturbations into a very simple form through the implementation of
canonical transformations and redefinitions of the lapse function
only, without recurring to any classical equations of motion
\cite{tens1}. Its consequences were explored in Ref.~\cite{tens2}.
However, tensor perturbations are very special (they are
automatically gauge invariant, their equations do not depend on the
matter background) and it remained to investigate if it would be
possible to do the same procedures to simplify the hamiltonian of
scalar and vector perturbations. Note that such perturbations are
not gauge invariant from the beginning, and they have contributions
from the matter perturbations, which renders the calculations much
more intricate.

The aim of this paper is to show that it is indeed possible to put
the complicated hamiltonians of scalar and vector perturbations of
Ref.~\cite{halli} into the very simple form of Ref.~\cite{MFB}
without using any classical background equations. We will exhibit
the canonical transformations and lapse functions redefinitions
which make the job. The simplified constraints obtained have direct
physical interpretations. The quantization of the theory yields a
very simple Wheeler-DeWitt equation for the perturbations and
background, which can be used with whatever interpretation and
choice of time one makes.

When the matter content is a perfect fluid, the Wheeler-DeWitt
equation assumes a Schr\"odinger form, and a further simplification
can be achieved provided one uses the ontological interpretation of
Bohm and de Broglie~\cite{dbb}. As in this case a quantum Bohmian
trajectory $a_q(t)$) at zeroth order can be defined, a time
dependent unitary transformation can be implemented in the scalar
perturbation sector using this $a_q(t)$, and, like in
Ref.~\cite{MFB}, the hamiltonian can be further simplified rendering
equations governing the scalar perturbations which are formally
equivalent to simple equations for a scalar field with an effective
mass depending on the quantum solution for the scale factor
$a_q(t)$, the quantum "pump field".

The paper is organized as follows. In the following section, we
specify the action and hamiltonian by restricting attention to the
particular case of a Friedmann-Lema\^{\i}tre-Robertson-Walker (FLRW)
background and perturbations around it, without yet making any
separation in scalar, vector and tensor perturbations. In sections
III and IV we analyze the cases of vector and scalar perturbations,
respectively. We concentrate in the hydrodynamical fluid case
letting the scalar field case to Appendix A. In section IV we
present all the steps to simplify the scalar part of the
hamiltonian. In section V we quantize this system. After separating
the background Schr\"odinger equation from the perturbed one, we
show how to use the Bohm-de Broglie interpretation in order to
perform the last canonical transformations which yields quantum
equations for the perturbations with the same form as those
presented in Ref.~\cite{MFB}. Finally, Sec. VI ends this paper with
some general conclusions. Appendix B presents the explicit canonical
transformations used in section IV.

\section{Linear Cosmological Perturbations}

Let the geometry of spacetime be given by
\begin{equation}
\label{perturb}
g_{\mu\nu}=g^{(0)}_{\mu\nu}+h_{\mu\nu},
\end{equation}
where $g^{(0)}_{\mu\nu}$ represents a homogeneous and isotropic
cosmological background,
\begin{equation}
\label{linha-fried}
ds^{2}=g^{(0)}_{\mu\nu}dx^{\mu}dx^{\nu}=N^{2}(t)dt^2-a^{2}(t)\gamma_{ij}dx^{i}dx^{j},
\end{equation}
where $\gamma_{ij}$ is the spatial metric of the spacelike
hypersurfaces with constant curvature $K=0,\pm 1$, and the
$h_{\mu\nu}$ are the linear perturbations, which we decompose into
\begin{eqnarray}
\label{perturb-componentes}
h_{00}&=&2N^{2}\phi ,\nonumber \\
h_{0i}&=&-NaA_{i} ,\\
h_{ij}&=&a^{2}\epsilon_{ij} .\nonumber
\end{eqnarray}
Substituting Eq.~(\ref{perturb-componentes}) into the Einstein-Hilbert action
\begin{equation}
\label{acao-grav}
S_{gr}=-\frac{1}{6l^{2}}\int d^{4}x\sqrt{-g}R,
\end{equation}
where $l^2=8\pi G/3$, yields the zeroth and second order actions
\begin{equation}
\label{acao-grav-bg} S^{(0)}_{gr}=\frac{1}{6l^{2}}\int
d^{4}xN\gamma^{\frac{1}{2}}a^{3}\biggl(-\frac{6\dot{a}^{2}}{a^{2}N^{2}}+\frac{6K}{a^{2}}\biggr)
\end{equation}
\begin{widetext}
\begin{eqnarray}
\label{acao-grav-2ordem} \delta_{2}S_{gr}&=&\frac{1}{6l^{2}}\int
d^{4}xN\gamma^{\frac{1}{2}}a^{3}\biggl[\frac{1}{4N^{2}}\dot{\epsilon^{ij}}\dot{\epsilon_{ij}}-
\frac{1}{4N^{2}}\dot{\epsilon}^{2}+\frac{1}{aN}\dot{A_{i}}\epsilon^{ij}\,_{|j}-\frac{1}{aN}\dot{\epsilon}A^{i}\,_{|i}+\frac{\dot{a}}{a^{2}N}(-4\phi
A^{i}\,_{|i}+2A_{i}\epsilon^{ij}\,_{j})\nonumber\\
&+&\frac{\dot{a}}{aN^{2}}(-\epsilon\dot{\epsilon}-2\dot{\epsilon}\phi+2\dot{\epsilon_{ij}}\epsilon^{ij})
+\frac{\dot{a}^{2}}{a^{2}N^{2}}(-3\epsilon\phi-9\phi^{2}+3A^{i}A_{i}-\frac{3}{4}\epsilon^{2}+\frac{3}{2}\epsilon^{ij}\epsilon_{ij})
+\frac{1}{a^{2}}A^{i|j}A_{[i|j]}-\frac{1}{4a^{2}}\epsilon_{ij|k}\epsilon^{ij|k}\nonumber
\\&+&\frac{1}{2a^{2}}\epsilon^{ij}\,_{|j}\epsilon_{i}\,^{k}\,_{|k}
+\frac{1}{a^{2}}\phi_{|i}\epsilon^{ij}\,_{|j}-\frac{1}{2a^{2}}\epsilon_{|i}\epsilon^{ij}\,_{|j}-\frac{1}{a^{2}}\phi_{|i}\epsilon^{|i}+
\frac{1}{4a^{2}}\epsilon_{|i}\epsilon^{|i}
+\frac{K}{a^{2}}(A^{i}A_{i}-\epsilon^{ij}\epsilon_{ij}-3\phi^{2}-\epsilon\phi+\frac{1}{4}\epsilon^{2})\biggr]
\end{eqnarray}
\end{widetext}
The first order action was discarded because we are assuming that the mean value of the perturbations over
the spatial sections are null:
\begin{equation}
\int d^{3}x \gamma^{\frac{1}{2}}h_{\mu\nu}=0.
\end{equation}

These are the actions for the gravitational sector. Let us now focus
on the action for the matter sector. We will concentrate on perfect
fluids for two reasons: 1) When quantizing the theory, a time
variable appears naturally putting the Wheeler-DeWitt equation into
a Schr\"odinger form. 2) The background quantum solutions of these
models are quite simple \cite{pinto,fabris}, contrary to the scalar
field case, where the quantum trajectories are implicity functions
of time \cite{pinto2}. We will let the discussion concerning the
scalar field to Appendix 1, which follows the same lines as below.

We will restrict the construction of the hamiltonian to the $K=0$ case and postpone the $K \neq 0$,
which is more intricate, to a future publication.

Following the approach of Ref.~\cite{MFB}, the lagrangian density of
the perfect fluid is
\begin{equation}
\label{lagrange-densidade-hidro}
\pounds_{m}=-\varepsilon,
\end{equation}
where
\begin{equation}
\label{densidade-energia-hidro}
\varepsilon=\rho[m_{0}+\Pi(p,\rho)],
\end{equation}
and
\begin{equation}
\label{potencial-hidro}
\Pi(p,\rho)=\int_{0}^{\rho}\frac{dp}{d\rho^{\prime}}\frac{d\rho^{\prime}}{\rho^{\prime}}-\frac{p}{\rho},
\end{equation}
where $\rho$ is the number density of particles, $m_{0}$ is their
rest mass, and $p$ is an arbitrary function of $\rho$, which will
be identified with the pressure. The particle number density
$\rho$ is given by
\begin{equation}
\label{densidade-particulas-hidro}
\rho=\frac{F(\mathbf{a}^{i})\sqrt{g_{\mu\nu}\frac{\partial
x^{\mu}}{\partial\sigma} \frac{\partial
x^{\nu}}{\partial\sigma}}}{\sqrt{-g}J} ,
\end{equation}
where $F$ is an arbitrary function of lagrangian variables, $\sigma$ is a time parameter along the
particle world lines, and $J$ is the jacobian of the transformation from lagrangian variable to
eulerian ones.

The energy-momentum tensor of the fluid reads
\begin{equation}
\label{momento-energia-hidro}
T^{\mu\nu}=-\frac{2}{\sqrt{-g}}\frac{\delta(\sqrt{-g}\pounds)}{\delta g_{\mu\nu}}=
\varepsilon V^{\mu}V^{\nu}-p(g^{\mu\nu}-V^{\mu}V^{\nu}),
\end{equation}
where it is clear that $\varepsilon$ e $p$  corresponds to the energy density and pressure, respectively.
The sound velocity $c_{s}$ is defined by
\begin{equation}
\label{velocidade-som} c_{s}^{2}=\frac{\partial\varepsilon}{\partial
p} .
\end{equation}

Perturbations displace the particles from their background positions
$x_{0}^{\mu}$ to the $x^{\mu}$ positions given by
\begin{equation}
\label{perturb-hidro} x^{\alpha}_{0}\rightarrow
x^{\alpha}=x^{\alpha}_{0}+\xi^{\alpha}(x_{0}) ,
\end{equation}
meaning a change into their eulerian position, which implies
modifications in the jacobian,
\begin{equation}
\label{jacobiano-hidro}
J=J_{0}\biggl(1+\dot{\xi^{0}}+\xi^{i}\,_{,i}+\frac{1}{2}\xi^{i}\,_{,i}\xi^{j}\,_{,j}+\dot{\xi^{0}}\xi^{i}\,_{,i}-
\frac{1}{2}\xi^{i}\,_{,j}\xi^{j}\,_{,i}-\xi^{0}\,_{,i}\dot{\xi^{i}}\biggr),
\end{equation}
in the determinant,
\begin{widetext}
\begin{eqnarray}
\label{determinante-hidro}
\sqrt{-g}(x_0+\xi)&=&\sqrt{-g^{(0)}}(x_0)\biggl(1+\phi-\frac{1}{2}\epsilon+\frac{\dot{N}}{N}\xi^{0}+
\frac{3\dot{a}}{a}\xi^{0}
-\frac{1}{2}\phi^{2}-\frac{1}{2}\epsilon\phi+\frac{1}{2}A^{i}A_{i}-\frac{1}{4}\epsilon^{ij}\epsilon_{ij}+
\frac{1}{8}\epsilon^{2}+\frac{\dot{N}}{N}\phi\xi_{0}\nonumber \\
&-&\frac{1}{2}\frac{\dot{N}}{N}\epsilon\xi^{0}+\frac{3\dot{a}}{a}\xi^{0}\phi-\frac{3}{2}
\frac{\dot{a}}{a}\epsilon\xi^{0} +\dot{\phi}\xi^{0}+\phi_{i}\xi^{i}
-\frac{1}{2}\dot{\epsilon}\xi^{0}+\frac{3\dot{N}\dot{a}}{Na}\xi^{0}\,^{2}+
\frac{3\dot{a}^{2}}{a^{2}}\xi^{0}\,^{2}
+\frac{\ddot{N}}{2N}\xi^{0}\,^{2}+\frac{3\ddot{a}}{2a}\xi^{0}\,^{2}
-\frac{1}{2}\epsilon_{,i}\xi^{i}\biggr) ,
\nonumber \\
&&
\end{eqnarray}
and in
\begin{eqnarray}
\label{raiz-gmini-fmi-fni-hidro} \sqrt{g_{\mu\nu}\frac{\partial
x^{\mu}}{\partial\sigma}\frac{\partial x^{\nu}}{\partial\sigma}}&=&
\sqrt{g^{(0)}_{\mu\nu}\frac{\partial
x^{\mu}_{0}}{\partial\sigma}\frac{\partial
x^{\nu}_{0}}{\partial\sigma}}
\biggl(1+\phi+\frac{\dot{N}}{N}\xi^{0}+\dot{\xi^{0}}
+\frac{\dot{N}}{N}\phi\xi^{0}+\dot{\phi}\xi^{0}+\phi_{|i}\xi^{i}+\frac{\ddot{N}}{2N}\xi_{0}^{2}+\phi\dot{\xi_{0}} \nonumber \\
&+&\frac{\dot{N}}{N}\dot{\xi^{0}}\xi^{0}-\frac{a}{N}A_{i}\dot{\xi^{i}}-\frac{1}{2}\frac{a^{2}}{N^{2}}\gamma_{ij}
\dot{\xi^{i}}\dot{\xi^{j}} -\frac{1}{2}\phi^{2}\biggr) .
\end{eqnarray}
The particle number density at point $x_0$ is then given by
\begin{eqnarray}
\label{perturb-densidade-particulas-hidro-0}
\rho(x_{0})&=&\rho_{0}\biggl(1+\frac{1}{2}\epsilon-\xi^{i}\,_{|i}+\dot{\xi^{i}}\,_{|i}\xi^{0}+\xi^{i}\,_{|j|i}\xi^{i}-\frac{a}{N}A_{i}\dot{\xi^{i}}
-\frac{1}{2}\frac{a^{2}}{N^{2}}\gamma_{ij}\dot{\xi^{i}}\dot{\xi^{j}}-\frac{1}{2}A_{i}A^{i}+\frac{1}{4}\epsilon_{ij}\epsilon^{ij}+\frac{1}{8}\epsilon^{2} \nonumber \\
&+&\xi^{0}_{|i}\dot{\xi^{i}}+\frac{1}{2}\xi^{i}\,_{|i}\xi^{j}\,_{|j}+\frac{1}{2}\xi^{i}\,_{|j}\xi^{j}\,_{|i}-\frac{1}{2}\epsilon\xi^{i}\,_{|i}\biggr)
\end{eqnarray}
Substituting all that in Eq.~(\ref{potencial-hidro}) and finally in
Eq.~(\ref{densidade-energia-hidro}), yields
\begin{eqnarray}
\label{perturb-acao-hidro}
\delta_{2}S_{m}&=&-\int d^{4}xNa^{3}\gamma^{\frac{1}{2}}\biggl[\varepsilon_{0}\biggl(-\frac{1}{2}\phi^{2}+\frac{1}{2}A^{i}A_{i}-\phi\xi^{i}\,_{|i}\biggr)
+p_{0}\biggl(\frac{1}{2}\epsilon\phi+\frac{1}{4}\epsilon^{ij}\epsilon_{ij}-\frac{1}{8}\epsilon^{2}-\phi\xi^{i}\,_{|i}\biggr) \nonumber \\
&-&\frac{1}{2}(\varepsilon_{0}+p_{0})\biggl(\frac{a^{2}}{N^{2}}\dot{\xi^{i}}\dot{\xi_{i}}+2\frac{a}{N}A_{i}\dot{\xi^{i}}+A_{i}A^{i}\biggr)
+\frac{1}{2}c_{s}^{2}(\varepsilon_{0}+p_{0})\biggl(\frac{1}{4}\varepsilon^{2}+\xi^{i}\,_{|i}\xi^{j}\,_{|j}-
\epsilon\xi^{i}\,_{|i}\biggr)\biggr] .
\end{eqnarray}
The total lagrangian including the gravitational sector then reads
\begin{eqnarray}
\label{lagrange-total-hidro}
L&=&-\frac{\dot{a}^{2}aV}{l^{2}N}-Na^{3}\varepsilon_{0}V
+\frac{Na}{6l^{2}}\int d^{3}x\gamma^{\frac{1}{2}}\biggl(A^{i|j}A_{[i|j]}-\frac{1}{4}\epsilon^{ij|k}\epsilon_{ij|k}+
\frac{a}{N}\dot{A_{i}}\epsilon^{ij}\,_{|j}+\frac{1}{2}\epsilon^{ij}\,_{|j}\epsilon_{i}\,^{k}\,_{|k}+\phi_{|i}\epsilon^{ij}\,_{|j} \nonumber \\
&-&\frac{1}{2}\epsilon_{|i}\epsilon^{ij}\,_{|j}-\phi_{|i}\epsilon^{|i}+\frac{1}{4}\epsilon_{|i}\epsilon^{|i}\biggr)+\frac{a^{3}}{24l^{2}N}
\int
d^{3}x\gamma^{\frac{1}{2}}\dot{\epsilon^{ij}}\dot{\epsilon_{ij}}
+\frac{a\dot{a}^{2}}{6l^{2}N}\int
d^{3}x\gamma^{\frac{1}{2}}\biggl(-9\phi^{2}- 3\epsilon\phi
-\frac{3}{4}\epsilon^{2}+3A^{i}A_{i}+\frac{3}{2}\epsilon^{ij}\epsilon_{ij}\biggr)\nonumber\\&-&\frac{2a\dot{a}}{3l^{2}}\int
d^{3}x\gamma^{\frac{1}{2}}\biggl(\phi A^{i}\,_{|i}
-\frac{1}{2}A_{i}\epsilon^{ij}\,_{|j}\biggr)-\frac{a^{3}}{24l^{2}N}\int
d^{3}x\gamma^{\frac{1}{2}}\dot{\epsilon}^{2}+\frac{a^{2}\dot{a}}{3l^{2}N}\int
d^{3}x\gamma^{\frac{1}{2}}\biggl(\epsilon^{ij}\dot{\epsilon_{ij}}-
\frac{1}{2}\epsilon\dot{\epsilon}-\phi\dot{\epsilon}\biggr)-\frac{a^{2}}{6l^{2}}\int
d^{3}x\gamma^{\frac{1}{2}}\dot{\epsilon}A^{i}\,_{|i}
\nonumber\\&-&Na^{3}\varepsilon_{0}\int
d^{3}x\gamma^{\frac{1}{2}}\biggl(-\frac{1}{2}\phi^{2}+\frac{1}{2}A^{i}A_{i}
-\phi\xi^{i}\,_{|i}\biggr)-Na^{3}p_{0}\int d^{3}x
\gamma^{\frac{1}{2}}\biggl(\frac{1}{2}\epsilon\phi+\frac{1}{4}\epsilon^{ij}\epsilon_{ij}-
\frac{1}{8}\epsilon^{2}-\phi\xi^{i}\,_{|i}\biggr) \nonumber \\
&+&\frac{1}{2}Na^{3}(\varepsilon_{0}+p_{0})\int
d^{3}x\gamma^{\frac{1}{2}}\biggl(\frac{a^{2}}{N^{2}}\dot{\xi^{i}}\dot{\xi_{i}}+
2\frac{a}{N}A_{i}\dot{\xi^{i}}+A_{i}A^{i}\biggr)
-\frac{1}{2}c_{s}^{2}Na^{3}(\varepsilon_{0}+p_{0})\int d^{3}x
\gamma^{\frac{1}{2}}\biggl(\frac{1}{4}\varepsilon^{2}+
\xi^{i}\,_{|i}\xi^{j}\,_{|j}-\epsilon\xi^{i}\,_{|i}\biggr) .\nonumber \\
&&
\end{eqnarray}

The procedure of Ref.~\cite{MFB} to simplify Eq.~(\ref{lagrange-total-hidro}) begins as follows:
using the background equation of motion
\begin{equation}
\label{eq-mov-friedp-hidro}
\ddot{a}=-\frac{\dot{a}^{2}}{2a}+\frac{\dot{a}\dot{N}}{N}-\frac{3l^{2}N^{2}a}{2}p_{0},
\end{equation}
and discarding a total time derivative
\begin{equation}
\label{superficie-hidro}
\biggl[\frac{a^{2}\dot{a}}{6l^{2}N}\int d^{3}x\gamma^{\frac{1}{2}}(\epsilon^{ij}\epsilon_{ij}-
\frac{1}{2}\epsilon^{2})\biggr]\dot{\,},
\end{equation}
we obtain

\begin{eqnarray}
\label{lagrange-c1-hidro}
L&=&-\frac{\dot{a}^{2}aV}{l^{2}N}-Na^{3}\varepsilon_{0}V+\frac{Na}{6l^{2}}\int d^{3}x\gamma^{\frac{1}{2}}\biggl(A^{i|j}A_{[i|j]}
-\frac{1}{4}\epsilon^{ij|k}\epsilon_{ij|k}+\frac{a}{N}\dot{A_{i}}\epsilon^{ij}\,_{|j}+\frac{1}{2}\epsilon^{ij}\,_{|j}\epsilon_{i}\,^{k}\,_{|k}+
\phi_{|i}\epsilon^{ij}\,_{|j} \nonumber \\
&-&\frac{1}{2}\epsilon_{|i}\epsilon^{ij}\,_{|j}-\phi_{i}\epsilon^{|i}+\frac{1}{4}\epsilon_{i}\epsilon^{i}\biggr)+\frac{a^{3}}{24l^{2}N}\int
d^{3}x \gamma^{\frac{1}{2}}\dot{\epsilon^{ij}}\dot{\epsilon_{ij}}+
\frac{a\dot{a}^{2}}{6l^{2}N}\int
d^{3}x\gamma^{\frac{1}{2}}(-9\phi^{2}- 3\epsilon\phi +3A^{i}A_{i})
\nonumber
\\&-&\frac{a^{3}}{24l^{2}N}\int
d^{3}x\gamma^{\frac{1}{2}}\dot{\epsilon}^{2}
-\frac{2a\dot{a}}{3l^{2}}\int d^{3}x\gamma^{\frac{1}{2}}\biggl(\phi
A^{i}\,_{|i}-\frac{1}{2}A_{i}\epsilon^{ij}\,_{|j}\biggr)-\frac{a^{2}\dot{a}}{3l^{2}N}\int
d^{3}x\gamma^{\frac{1}{2}}\phi\dot{\epsilon}-
\frac{a^{2}}{6l^{2}}\int d^{3}x\gamma^{\frac{1}{2}}\dot{\epsilon}A^{i}\,_{|i} \nonumber \\
&-&Na^{3}\varepsilon_{0}\int d^{3}x\gamma^{\frac{1}{2}}\biggl(-\frac{1}{2}\phi^{2}+\frac{1}{2}A^{i}A_{i}-\phi\xi^{i}\,_{|i}\biggr)
-Na^{3}p_{0}\int d^{3}x \gamma^{\frac{1}{2}}\biggl(\frac{1}{2}\epsilon\phi-\phi\xi^{i}\,_{|i}\biggr) \nonumber \\
&+&\frac{1}{2}Na^{3}(\varepsilon_{0}+ p_{0})\int
d^{3}x\gamma^{\frac{1}{2}}\biggl(\frac{a^{2}}{N^{2}}\dot{\xi^{i}}\dot{\xi_{i}}+2\frac{a}{N}A_{i}\dot{\xi^{i}}+A_{i}A^{i}\biggr)
-\frac{1}{2}c_{s}^{2}Na^{3}(\varepsilon_{0}+p_{0})\int d^{3}x
\gamma^{\frac{1}{2}}\biggl(\frac{1}{4}\varepsilon^{2}+\xi^{i}\,_{|i}\xi^{j}\,_{|j}-
\epsilon\xi^{i}\,_{|i}\biggr) \nonumber \\
&&
\end{eqnarray}

Now using the other background equation
\begin{equation}
\label{eq-mov-friede-hidro}
\frac{{\dot{a}}^2 a}{6l^{2}N}=\frac{Na^{3}\epsilon_{0}}{6},
\end{equation}
we arrive at the simplified lagrangian
\begin{eqnarray}
\label{lagrange-c2-hidro}
L&=&-\frac{\dot{a}^{2}aV}{l^{2}N}-Na^{3}\varepsilon_{0}V+\frac{Na}{6l^{2}}\int d^{3}x\gamma^{\frac{1}{2}}\biggl(A^{i|j}A_{[i|j]}
-\frac{1}{4}\epsilon^{ij|k}\epsilon_{ij|k}+\frac{a}{N}\dot{A_{i}}\epsilon^{ij}\,_{|j}+\frac{1}{2}\epsilon^{ij}\,_{|j}\epsilon_{i}\,^{k}\,_{|k}+\phi_{|i}\epsilon^{ij}\,_{|j} \nonumber \\
&-&\frac{1}{2}\epsilon_{|i}\epsilon^{ij}\,_{|j}-\phi_{|i}\epsilon^{|i}+\frac{1}{4}\epsilon_{|i}\epsilon^{|i}\biggr)+
\frac{a^{3}}{24l^{2}N}\int d^{3}x\gamma^{\frac{1}{2}}\dot{\epsilon^{ij}}\dot{\epsilon_{ij}}
-\frac{a^{3}}{24l^{2}N}\int d^{3}x\gamma^{\frac{1}{2}}\dot{\epsilon}^{2}-\frac{a\dot{a}^{2}}{l^{2}N}\int d^{3}x\gamma^{\frac{1}{2}}\phi^{2} \nonumber \\
&-&\frac{2a\dot{a}}{3l^{2}}\int
d^{3}x\gamma^{\frac{1}{2}}\biggl(\phi
A^{i}\,_{|i}-\frac{1}{2}A_{i}\epsilon^{ij}\,_{|j}\biggr)
-\frac{a^{2}\dot{a}}{3l^{2}N}\int
d^{3}x\gamma^{\frac{1}{2}}\phi\dot{\epsilon}-\frac{a^{2}}{6l^{2}}\int
d^{3}x\gamma^{\frac{1}{2}}\dot{\epsilon}A^{i}\,_{|i}
-Na^{3}(\varepsilon_{0}+p_{0})\int d^{3}x \gamma^{\frac{1}{2}}\biggl(\frac{1}{2}\epsilon\phi-\phi\xi^{i}\,_{|i}\biggr) \nonumber \\
&+&\frac{1}{2}Na^{3}(\varepsilon_{0}+p_{0})\int
d^{3}x\gamma^{\frac{1}{2}}\biggl(\frac{a^{2}}{N^{2}}\dot{\xi^{i}}\dot{\xi_{i}}+
2\frac{a}{N}A_{i}\dot{\xi^{i}}+A_{i}A^{i}\biggr)
-\frac{1}{2}c_{s}^{2}Na^{3}(\varepsilon_{0}+p_{0})\int d^{3}x
\gamma^{\frac{1}{2}}\biggl(\frac{1}{4}\varepsilon^{2}+
\xi^{i}\,_{|i}\xi^{j}\,_{|j}-\epsilon\xi^{i}\,_{|i}\biggr).\nonumber \\
&&
\end{eqnarray}

Note that it is not necessary to use Eq.~(\ref{eq-mov-friede-hidro}) in order to pass from
Eq.~(\ref{lagrange-c1-hidro}) to Eq.~(\ref{lagrange-c2-hidro}): the redefinition of the lapse
function
\begin{equation}
\label{n-redef-hidro} N=:\tilde{N}\biggl[1+\frac{1}{2V}\int
d^{3}x\gamma^{\frac{1}{2}}(\epsilon\phi+\phi^{2}-A^{i}A_{i})\biggr]
\end{equation}
takes Eq.~(\ref{lagrange-c1-hidro}) into
Eq.~(\ref{lagrange-c2-hidro}). Note that these two lapse functions
related by Eq.~(\ref{n-redef-hidro}) are equivalent at first
order. Hence, this procedure does not modify the equations of
motion at first order wnen we make a time gauge choice.

Let us now calculate the hamiltonians of these lagrangians for perfect fluids with equation of state
\begin{equation}
\label{adiabatico}
p_{0}=\lambda\varepsilon_{0}
\end{equation}
The hamiltonian from Eq.~(\ref{lagrange-total-hidro}) reads
\begin{eqnarray}
\label{h-total-hidro}
H_{T}&=&-\frac{Nl^{2}P_{a}^{2}}{4aV}+N\frac{P_T}{a^{3\lambda}}+
\frac{Nl^{2}P_{a}^{2}}{aV^{2}}\int
d^{3}x\gamma^{\frac{1}{2}}\biggl(\frac{1}{8}\phi^{2}+\frac{1}{24}\epsilon\phi
-\frac{1}{8}A^{i}A_{i}-\frac{1}{32}\epsilon^{2}+\frac{5}{48}\epsilon^{ij}\epsilon_{ij}\biggr)+
\frac{NP_{a}}{6V}\int d^{x}\gamma^{\frac{1}{2}}\biggl(\phi
A^{i}\,_{|i}+\epsilon^{ij}\,_{|j}A_{i}\biggr)
\nonumber\\&+&\frac{2Nl^{2}P_{a}}{aV}\int
d^{3}x\pi^{ij}\epsilon_{ij}-\frac{Nl^{2}P_{a}}{2a^{2}V^{2}}\int
d^{3}x\epsilon\pi +\frac{NP_{a}}{12V}\int
d^{3}x\gamma^{\frac{1}{2}}\epsilon
A^{i}\,_{|i}+\frac{Nl^{2}P_{a}}{a^{2}V}\int d^{3}x\pi\phi
+\frac{6Nl^{2}}{a^{3}}\int
d^{3}x\frac{\pi^{ij}\pi_{ij}}{\gamma^{\frac{1}{2}}}\nonumber
\\&-&\frac{3l^{2}N}{a^{3}}\int
d^{3}x\frac{\pi^{2}}{\gamma^{\frac{1}{2}}} -\frac{Na}{4l^{2}}\int
d^{3}x\gamma^{\frac{1}{2}}A^{i}\,_{|i}A^{j}\,_{|j}-\frac{N}{a}\int
d^{3}x\pi A^{i}\,_{|i}
+\frac{N}{2a^{5}(\lambda+1)\varepsilon_{0}}\int
d^{3}x\frac{\pi_{\xi}^{i}\pi_{\xi}\,_{i}}{\gamma^{\frac{1}{2}}}-
\frac{N}{a}\int d^{3}x\pi_{\xi}^{i}A_{i} \nonumber \\
&-&\frac{Na}{6l^{2}}\int
d^{3}x\gamma^{\frac{1}{2}}\biggl(A^{i|j}A_{[i|j]}-
\frac{1}{4}\epsilon^{ij|k}\epsilon_{ij|k}+\frac{1}{2}\epsilon^{ij}\,_{|j}\epsilon_{i}\,^{k}\,_{|k}
+\phi_{|i}\epsilon^{ij}\,_{|j}-\frac{1}{2}\epsilon_{|i}\epsilon^{ij}\,_{|j}-\phi_{|i}\epsilon^{|i}+\frac{1}{4}\epsilon_{|i}\epsilon^{|i}\biggr) \nonumber \\
&+&Na^{3}\varepsilon_{0}\int
d^{3}x\gamma^{\frac{1}{2}}\biggl(-\frac{1}{2}\phi^{2}+\frac{1}{2}A^{i}A_{i}-\phi\xi^{i}\,_{|i}\biggr)
+Na^{3}\lambda\varepsilon_{0}\int
d^{3}x\gamma^{\frac{1}{2}}\biggl(\frac{1}{2}\epsilon\phi+
\frac{1}{4}\epsilon^{ij}\epsilon_{ij}-\frac{1}{8}\epsilon^{2}-\phi\xi^{i}\,_{|i}\biggr)
\nonumber\\&+&\frac{1}{2}Na^{3}\lambda(\lambda+1)\varepsilon_{0}\int
d^{3}x\gamma^{\frac{1}{2}}\biggl(\frac{1}{4}
\epsilon^{2}+\xi^{i}\,_{|i}\xi^{j}\,_{|j}-\epsilon\xi^{i}\,_{|i}\biggr)
,
\end{eqnarray}
while that from Eq.~(\ref{lagrange-c1-hidro}) is given by
\begin{eqnarray}
\label{h-c1-hidro}
H_{T}&=&-\frac{Nl^{2}P_{a}^{2}}{4aV}+N\frac{P_T}{a^{3\lambda}}+\frac{Nl^{2}P_{a}^{2}}{aV^{2}}\int
d^{3}x\gamma^{\frac{1}{2}}\biggl(\frac{1}{8}\phi^{2}+
\frac{1}{8}\epsilon\phi-\frac{1}{8}A^{i}A_{i}\biggr)+\frac{NP_{a}}{6V}\int
d^3{x}\gamma^{\frac{1}{2}}\biggl(\phi A^{i}\,_{|i}
+\epsilon^{ij}\,_{|j}A_{i}\biggr)+\frac{Nl^{2}P_{a}}{a^{2}V}\int d^{3}x\pi\phi \nonumber \\
&+&\frac{6Nl^{2}}{a^{3}}\int
d^{3}x\frac{\pi^{ij}\pi_{ij}}{\gamma^{\frac{1}{2}}}-
\frac{3l^{2}N}{a^{3}}\int d^{3}x\frac{\pi^{2}}{\gamma^{\frac{1}{2}}}
-\frac{Na}{4l^{2}}\int
d^{3}x\gamma^{\frac{1}{2}}A^{i}\,_{|i}A^{j}\,_{|j}-\frac{N}{a}\int
d^{3}x\pi A^{i}\,_{|i}
+\frac{N}{2a^{5}(\lambda+1)\varepsilon_{0}}\int
d^{3}x\frac{\pi_{\xi}^{i}\pi_{\xi}\,_{i}}{\gamma^{\frac{1}{2}}}\nonumber \\
&-& \frac{N}{a}\int d^{3}x\pi_{\xi}^{i}A_{i} -\frac{Na}{6l^{2}}\int
d^{3}x\gamma^{\frac{1}{2}}\biggl(A^{i|j}A_{[i|j]}-\frac{1}{4}\epsilon^{ij|k}\epsilon_{ij|k}+
\frac{1}{2}\epsilon^{ij}\,_{|j}\epsilon_{i}\,^{k}\,_{|k}
+\phi_{|i}\epsilon^{ij}\,_{|j}-\frac{1}{2}\epsilon_{|i}\epsilon^{ij}\,_{|j}-\phi_{|i}\epsilon^{|i}+\frac{1}{4}\epsilon_{|i}\epsilon^{|i}\biggr)
\nonumber \\
&+&Na^{3}\varepsilon_{0}\int
d^{3}x\gamma^{\frac{1}{2}}\biggl(-\frac{1}{2}\phi^{2}+\frac{1}{2}A^{i}A_{i}-\phi\xi^{i}\,_{|i}\biggr))
+Na^{3}\lambda\varepsilon_{0}\int
d^{3}x\gamma^{\frac{1}{2}}\biggl(\frac{1}{2}\epsilon\phi-\phi\xi^{i}\,_{|i}\biggr)
\nonumber \\
&+&\frac{1}{2}Na^{3}\lambda(\lambda+1)\varepsilon_{0}\int
d^{3}x\gamma^{\frac{1}{2}}
\biggl(\frac{1}{4}\epsilon^{2}+\xi^{i}\,_{|i}\xi^{j}\,_{|j}-\epsilon\xi^{i}\,_{|i}\biggr)
\end{eqnarray}
\end{widetext}
The quantity $P_T$ appearing in the second term of the zeroth order
term of both hamiltonians is just the kinematical constant
$P_T\equiv\varepsilon_0 a^{3\lambda+3} V$. We have introduced it as
a canonical momentum to a variable $T$ which is cyclic, implying
indeed that $P_T$ is a constant. We have made an inverse Routh
procedure. The variable $T$ plays the role of time when the system
is quantized. This form of the zeroth order hamltonian appears in
other approaches to a lagrangian formulation of fluids; see e.g.
Ref.~\cite{fluid} for details.

One can now use the total time derivative (\ref{superficie-hidro})
to construct the generator of canonical transformations
\begin{equation}
\label{gerador1-hidro} \mathcal{F}=a\tilde{P_{a}}-\int
d^{3}x\pi^{ij}\tilde{\epsilon_{ij}}-\frac{\tilde{P_{a}}a}{12V}
\int
d^{3}x\gamma^{\frac{1}{2}}(\tilde{\epsilon}^{ij}\tilde{\epsilon}_{ij}-\frac{1}{2}\tilde{\epsilon}^{2}),
\end{equation}
yielding
\begin{eqnarray}
\label{transf1-hidro}
a&=& \tilde{a}\biggl[1+\frac{1}{12V}
\int d^{3}x\gamma^{\frac{1}{2}}\biggl(\epsilon^{ij}\epsilon_{ij}-\frac{1}{2}\epsilon^{2}\biggr)\biggr]\nonumber \\
P_{a}&=& {\tilde{P}}_a\biggl[1-\frac{1}{12V}
\int d^{3}x\gamma^{\frac{1}{2}}\biggl(\epsilon^{ij}\epsilon_{ij}-\frac{1}{2}\epsilon^{2}\biggr)\biggr]\nonumber \\
{\tilde{\pi}}^{ij}&=& \pi^{ij}+\frac{aP_a}{6V}
\gamma^{\frac{1}{2}}\biggl(\epsilon^{ij}\epsilon_{ij}-\frac{1}{2}\epsilon^{2}\biggr)\nonumber\\
{\tilde{\epsilon}}^{ij}&=&\epsilon ^{ij}.
\end{eqnarray}
Using the fact that $\rho\propto a^{-3}$, the particle number density transforms to
\begin{equation}
\label{densidade-particulas-transf1}
\rho=\tilde{\rho}\biggl[1-\frac{1}{4V}\int
d^{3}x\gamma^{\frac{1}{2}}(\tilde{\epsilon^{ij}}\tilde{\epsilon_{ij}}-\frac{1}{2}\tilde{\epsilon^{2}})\biggr]=:\tilde{\rho}-\delta\rho
\end{equation}
Substituting this last equation into Eqs.~(\ref{potencial-hidro})
and (\ref{densidade-energia-hidro}) we obtain
\begin{equation}
\label{densidade-energia-transf1}
\varepsilon_{0}=\tilde{\varepsilon_{0}}-\frac{(\tilde{\varepsilon_{0}}+\tilde{p_{0}})}{\tilde{\rho}}\delta\rho.
\end{equation}
Inserting Eqs.~(\ref{transf1-hidro}) and
(\ref{densidade-energia-transf1}) into Eq.~(\ref{h-total-hidro}), we
obtain (\ref{h-c1-hidro}). Hence, in the lagrangian point of view,
one can pass from Eq.~(\ref{lagrange-total-hidro}) to
Eq.~(\ref{lagrange-c1-hidro}) without using any background equations
of motion. As we have shown that we can pass from
Eq.~(\ref{lagrange-c1-hidro}) to Eq.~(\ref{lagrange-c2-hidro}) just
through a redefinition of $N$, then it is proven that lagrangian
(\ref{lagrange-total-hidro}) is equivalent to lagrangian
(\ref{lagrange-c2-hidro}) at this order of approximation
irrespective of the classical background equations of motion.

In order to proceed from this
point\footnote{Equation~(\ref{lagrange-c2-hidro}) corresponds to
Eq.~(10.37) of Ref.~\cite{MFB} if one is restricted to scalar
perturbations, and if one reads $\beta$ in the latter as
$\beta=3a^2l^2(\epsilon_0+p_0)/2$.}, we will now separate the
perturbations into scalar, vector, and tensor perturbations. We
make the decomposition:
\begin{eqnarray}
\label{perturb-modos}
A_{i}&=&B_{|i}+S_{i} \nonumber \\
\epsilon_{ij}&=&2\psi\gamma_{ij}-2E_{|i|j}-F_{i|j}-F_{j|i}+w_{ij}
\end{eqnarray}
in the gravitational sector,
while the quantities $w_{ij}$, $F_{i}$ e $S_{i}$  satisfy
\begin{eqnarray}
\label{perturb-condicoes-esc}
S^{i}\,_{|i}&=&F^{i}\,_{|i}=0 \nonumber \\
w^{ij}\,_{|j}&=&0 \nonumber \\
w^{i}\,_{i}&=&0,
\end{eqnarray}
and
\begin{equation}
\label{perturb-decomp-hidro}
\xi^{i}=\eta^{i}+\zeta^{|i},
\end{equation}
with
\begin{equation}
\label{perturb-condicao-hidro} \eta^{i}\,_{|i}=0,
\end{equation}
in the matter sector. Substituting the above decompositions into
eq.~(\ref{lagrange-c2-hidro}) leads to a separation of this
lagrangian, with some total derivatives discarded, into three
independent sectors: scalar, vector and tensor sectors. We will
focus our attention on the vector and scalar sectors because the
case of tensor perturbations has already been treated in
Ref.~\cite{tens1}.

\section{Vector Perturbations}

Combining the contributions of gravitational and matter sectors and defining the gauge invariant
quantities
\begin{equation}
\label{perturb--invariante0-hidro}
V^{i}=S^i - \frac{a}{N}{\dot{F}}^i,
\end{equation}
and
\begin{equation}
\label{perturb--invariante-hidro}
\eta^{i}\,^{(gi)}=\eta^{i}+F^{i},
\end{equation}
we obtain
\begin{widetext}
\begin{equation}
\label{lagrange-vetor-hidro} L^{V}=\frac{Na}{12l^{2}}\int
d^{3}x\gamma^{\frac{1}{2}}V^{i|j}V_{i|j}
+\frac{1}{2}Na^{3}(\lambda+1)\varepsilon_{0}\int
d^{3}x\gamma^{\frac{1}{2}}
\biggl(\frac{a}{N}\dot{\eta}^{i}\,^{(gi)}+V^{i}\biggr)\biggl(\frac{a}{N}\dot{\eta}_{i}^{(gi)}+V_{i}\biggr).
\end{equation}

When constructing the hamiltonian, the primary constraint $\Pi_{i}\approx 0$ appears, where $\approx$ means
a weak equality in the sense of Dirac \cite{constrained}, and $\Pi_i$ is the momentum canonically conjugate
to $V^i$. The hamiltonian then reads

\begin{equation}
\label{h-escalares-hidro}
H^{V}=N\biggl[-\frac{l^{2}P_{a}^{2}}{4aV}+\frac{P_{T}}{a^{3\lambda}}+\int
d^{3}x \biggl(\frac{P^{i}P_i}{P_T a^{2-3\lambda}\gamma^{1/2}} +
\frac{V^i
P_i}{a}-\frac{a}{12l^{2}}\gamma^{\frac{1}{2}}V^{i|j}V_{i|j}\biggr)\biggr]+\int
d^{3}x\Lambda^{i}\pi_{i},
\end{equation}
where $P_i$ is the momentum canonically conjugate to
$\eta^{i}\,^{(gi)}$, and the $\Lambda^i$ are Lagrange multipliers.
\end{widetext}

The conservation of $\Pi_{i}$ in time imposes a secondary constraint

\begin{equation}
\label{fi3-vetor-hidro} \dot{\Pi^i}=\{\Pi^i,H^V\}=\phi_{2}^{i}=
\frac{1}{a}\pi_{\eta}^{i}-\frac{a}{6l^{2}}\gamma^{\frac{1}{2}}V^{i|j}\,_{|j}.
\end{equation}
The conservation in time of $\phi_2^i$ fixes the lagrange multipliers $\Lambda^i$ to the value
\begin{equation}
\label{lambdav-hidro} \Lambda^{i}=\frac{l^{2}P_{a}}{a^{2}V}V^{i}.
\end{equation}
Then the equations of motion for $V^i$ and $\eta^{i}\,^{(gi)}$ imply that

\begin{equation}
\label{vetor-hidro}
V^{i}=\frac{V^{i}_{0}}{a^{2}},
\end{equation}
and
\begin{equation}
\label{fi-vetor-hidro}
\varphi^{i}\equiv\biggl(\frac{a}{N}\dot{\eta}^{i}\,^{(gi)}+V^{i}\biggr)=
\frac{\nabla^2 V^{i}_{0}}{(\lambda+1)P_T a^{1-\lambda}}
\end{equation}
These solutions correspond to the classical result, which was
obtained without recurring to the classical background equations.

The two constraints obtained are second class. After defining the
corresponding Dirac brackets \cite{constrained}, they become strong
equalities which can be used to obtain some variables from others.

\section{Scalar Perturbations}

Defining the quantities

\begin{equation}
\label{escalares0-invariante-hidro}
F=B-\frac{a}{N}\dot{E},
\end{equation}
\begin{equation}
\label{escalares-invariante-hidro}
\zeta^{(gi)}=\zeta+E,
\end{equation}
which is a gauge invariant quantity, and
\begin{equation}
\label{def-fi-escalares-hidro}
\varphi=\frac{\sqrt{6}la^{2}\sqrt{(\lambda+1)\varepsilon_{0}}}{\sqrt{\lambda}}(\frac{a}{N}\dot{\zeta}^{(gi)}+F),
\end{equation}
which can be identified with the perturbed velocity potential of the fluid particles,
the scalar lagrangian reads
\begin{widetext}
\begin{eqnarray}
\label{lagrange-escalares-fi-hidro}
L^{E}&=&\frac{Na}{3l^{2}}\int d^{3}x \gamma^{\frac{1}{2}}(\psi^{,i}\psi_{,i}-2\phi^{,i}\psi_{,i})
-\frac{2a^{2}}{3l^{2}}\int d^{3}x \gamma^{\frac{1}{2}}(\dot{\psi}+\frac{\dot{a}}{a}\phi)F^{,i}\,_{,i}
-\frac{a^{3}}{Nl^{2}}\int d^{3}x\gamma^{\frac{1}{2}}(\dot{\psi}+\frac{\dot{a}}{a}\phi)^{2} \nonumber \\
&-&\frac{Na^{3}(\lambda+1)\varepsilon_{0}}{2}\int
d^{3}x\gamma^{\frac{1}{2}}
[\lambda(3\psi-\zeta^{,i}\,^{(gi)}_{,i})^{2}+2\phi(3\psi-\zeta^{,i}\,^{(gi)}_{,i})]
+\frac{N\lambda}{12l^{2}a} \int
d^{3}x\gamma^{\frac{1}{2}}\varphi^{,i}\varphi_{,i}.
\end{eqnarray}

As in the vector sector, some constraints appear, and, because of
definition (\ref{def-fi-escalares-hidro}) which involves a time
derivative, we have to use the Ostrogradsky method \cite{ostro}
through the definition $\pi_{\zeta}=\partial
L/\partial\dot{\zeta}-{\dot{\pi}}_{\varphi}$. The constraints are
\begin{equation}
\label{vinculos-escalares-hidro} \phi_{1}=P_{N} ;\;
\phi_{2}=\pi_{F};\; \phi_{3}=\pi_{\phi} ;\; \phi_{7}=\pi_{\varphi}
;\; \phi_{9}=P_{\mu} ,
\end{equation}
and the hamiltonian is
\begin{equation}
\label{h-escalares-hidro} H=N\biggl[H_{0}+\int
d^{3}x\Lambda_{\phi}\pi_{\phi}+\int d^{3}x\Lambda_{F}\pi_{F} +\int
d^{3}x\Lambda_{\varphi}\pi_{\varphi}\biggr] + \Lambda_N P_N
\end{equation}
where $H_{0}$ reads
\begin{eqnarray}
\label{h0-escalares-hidro}
H_{0}&=&-\frac{l^{2}P_{a}^{2}}{4aV}+\frac{P_{T}}{a^{3\lambda}}+\frac{(\lambda+1)P_{T}}{2a^{3\lambda}
V}\int d^{3}x\gamma^{\frac{1}{2}}
\biggl[\lambda(3\psi-\zeta^{,i}\,^{(gi)}_{,i})^{2}+2\phi(3\psi-\zeta^{,i}\,^{(gi)}_{,i})\biggr]+\frac{l^{2}P_{a}}{2a^{2}V}\int
d^{3}x\phi\pi_{\psi} \nonumber \\ &+&\frac{1}{a}\int
d^{3}x\pi_{\zeta}\biggl(\frac{\sqrt{V}\sqrt{\lambda}}{\sqrt{6}l\sqrt{(\lambda+1)P_{T}}}a^{-\frac{1}{2}(1-3\lambda)}\varphi-F\biggr)
-\frac{a}{l^{2}}\int
d^{3}x\gamma^{\frac{1}{2}}\biggl(\frac{l^{2}}{2a^{2}\gamma^{\frac{1}{2}}}\pi_{\psi}+\frac{1}{3}F^{,i}\,_{,i}{\biggr)}^{2}
\nonumber \\
&+&\frac{\lambda}{12l^{2}a}\int
d^{3}x\gamma^{\frac{1}{2}}\varphi\varphi^{,i}\,_{,i}\frac{a}{3l^{2}}
\int d^{3}x\gamma^{\frac{1}{2}}(\psi-2\phi)\psi^{,i}\,_{,i} .
\end{eqnarray}
\end{widetext}

Conservation in time of the primary constraints
(\ref{vinculos-escalares-hidro}) leads to the secondary
constraints
\begin{eqnarray}
\label{vinculos-secund-escalares-hidro}
\phi_{4}&=&\frac{H}{N} \nonumber \\
\phi_{5}&=&\frac{1}{a}\pi_{\zeta}+\frac{2a}{3l^{2}}\gamma^{\frac{1}{2}}\biggl(\frac{l^{2}}{2a^{2}
\gamma^{\frac{1}{2}}}\pi_{\psi}+\frac{1}{3}F^{,i}\,_{,i}{\biggr)}^{,j}\,_{,j}, \nonumber \\
\phi_{6}&=&-\frac{(\lambda+1)P_{T}}{a^{3\lambda}V}\gamma^{\frac{1}{2}}(3\psi-\zeta^{,i}\,_{,i})-
\frac{l^{2}P_{a}}{2a^{2}V}\pi_{\psi}+\frac{2a}{3l^{2}}\gamma^{\frac{1}{2}}\psi^{,i}\,_{,i}, \nonumber \\
\phi_{8}&=&\frac{1}{a}\frac{\sqrt{V}}{\sqrt{6}l\sqrt{(\lambda+1)P_{T}}}a^{-\frac{1}{2}(1-3\lambda)}
\pi_{\zeta}+\frac{\sqrt{\lambda}}{6l^{2}a}\gamma^{\frac{1}{2}}\varphi^{,i}\,_{,i}. \nonumber \\
&&
\end{eqnarray}
Neglecting third order terms, conservations in time of $\phi_{4}$
and $\phi_{6}$ are identically satisfied, whereas conservation of
$\phi_{5}$ and $\phi_{8}$ determines the lagrange multipliers
$\Lambda_{F}$ and $\Lambda_{\varphi}$. The Lagrange multiplier
$\Lambda_{F}$ reads\footnote{The explicit value of
$\Lambda_{\varphi}$ is not important for what follows.}
\begin{equation}
\label{lambdaf-hidro}
a\Lambda_{F}=\psi-\phi+\frac{l^{2}P_{a}}{aV}F.
\end{equation}
As $\Lambda_{F}=\dot{F}/N$, then
\begin{equation}
\frac{a}{N}\dot{F}=\psi-\phi-\frac{2\dot{a}}{N}F,
\end{equation}
which, when expressed in terms of the gauge invariant Bardeen
potentials, yields
\begin{equation}
\label{fi-psi-hidro}
\Phi=\Psi,
\end{equation}
a well known result.

The Poisson brackets among the constraints read
\begin{eqnarray}
\label{parenteses-poisson-hidro}
&&\{\phi_{2}(x),\phi_{5}(x')\}=-\frac{2a}{9l^{2}}\gamma^{\frac{1}{2}}\delta(x-x')^{,ij}\,_{,ij} \nonumber \\
&&\{\phi_{7}(x),\phi_{8}(x')\}=-\frac{\sqrt{\lambda}}{6l^{2}a}\gamma^{\frac{1}{2}}\delta(x-x')^{,i}\,_{,i} \nonumber \\
&&\{\phi_{6}(x),\phi_{5}(x')\}=\frac{2}{9l^{2}}\gamma^{\frac{1}{2}}\delta(x-x')^{,ij}\,_{,ij}
-\frac{l^2}{a^2V}\pi_{\psi}(x)\pi_{\zeta}(x')\nonumber \\
&&\{\phi_{6}(x),\phi_{8}(x')\}=\frac{\sqrt{(\lambda+1)P_{T}}}{\sqrt{6}l\sqrt{V}}a^{-\frac{3}{2}(1+\lambda)}
\gamma^{\frac{1}{2}}\delta(x-x')^{,i}\,_{,i} \nonumber \\
&& -\frac{l}{4\sqrt{6}\sqrt{V(\lambda+1)P_{T}}}(1-3\lambda)a^{-\frac{3}{2}(1+3\lambda)}
\pi_{\psi}(x)\pi_{\zeta}(x').\nonumber\\
\end{eqnarray}
Defining
\begin{equation}
\label{fi6-primeira-classe-hidro}
\bar{\phi_{6}}=\phi_{6}+\frac{1}{a}\phi_{2}+\frac{\sqrt{(\lambda+1)P_{T}}\sqrt{6}l}{\sqrt{\lambda}\sqrt{V}}a^{-\frac{1}{2}(1+3\lambda)}\phi_{7},
\end{equation}
one can prove that $\bar{\phi_{6}}$ is a first class constraint:
it has zero Poisson brackets with all others constraints up to
third order. We are then left with four second class constraints,
$\phi_{2}$, $\phi_{5}$, $\phi_{7}$ and $\phi_{8}$, Hence, from the
$10$ degrees of freedom of phase space corresponding to the
variables $\phi$, $\psi$, F, $\varphi$ and $\xi$, we have to
extract $4$ from the second class constraints and $2$x$2=4$ from
the two first class constraints $\bar{\phi_{6}}$ and $\phi_3$,
remaining $2$ degrees of freedom in phase space, as expected for
this problem.

In order to eliminate the second class constraints, we have to define the Dirac brackets associated with them.
The Dirac brackets among the variables of phase space which are not canonical are (excepting the ones involving
$F$ and $\pi_F$, which are not relevant)
\begin{eqnarray}
\label{parenteses-dirac-escalares-hidro}
\{\zeta^{,i}\,^{(gi)}_{,i}(x),\varphi(x')\}^{D}&=&-\frac{\sqrt{6}l\sqrt{V}a^{-\frac{1}{2}(1-3\lambda)}}{\sqrt{\lambda}
\sqrt{(\lambda+1)P_{T}}\gamma^{\frac{1}{2}}}\delta(x-x')\nonumber \\
\{P_{a},\varphi(x)\}^{D}&=&\frac{1}{2a}(1-3\lambda)\varphi(x)
\end{eqnarray}

Defining the quantities
\begin{eqnarray}
\label{canonicas-escalares-hidro}
\varphi_{(c)}&=:&a^{\frac{1}{2}(1-3\lambda)}\varphi, \nonumber \\
\pi_{\varphi}\,_{(c)}&=:&-\frac{\sqrt{\lambda}\sqrt{(\lambda+1)P_{T}}}{\sqrt{6}l
\sqrt{V}}\gamma^{\frac{1}{2}}(3\psi-\zeta^{,i}\,^{(gi)}_{,i}) , \nonumber \\
\pi_{\psi}\,_{(c)}&=&\pi_{\psi}-\frac{3\sqrt{\lambda}\sqrt{(\lambda+1)P_{T}}}{\sqrt{6}l\sqrt{V}}
\gamma^{\frac{1}{2}}\varphi_{(c)},
\end{eqnarray}
we obtain that the Dirac brackets for these quantities are canonical. The hamiltonian in terms of
these new variables then reads
\begin{equation}
\label{h-vinculos-escalares-hidro}
H=NH_{0}-N\int d^{3}x\phi\phi_{6}+\int d^{3}x\Lambda_{\phi}\pi_{\phi}
\end{equation}
where $H_{0}$ is given by
\begin{widetext}
\begin{eqnarray}
\label{h0-canonica-escalares-hidro}
H_{0}&=&-\frac{l^{2}P_{a}^{2}}{4aV}+\frac{P_{T}}{a^{3\lambda}}+\frac{3l^{2}}{a^{3\lambda}}\int
d^{3}x \frac{\pi_{\varphi}^{2}\,_{(c)}}{\gamma^{\frac{1}{2}}}
-\frac{\lambda}{12l^{2}a^{2-3\lambda)}}\int
d^{3}x\gamma^{\frac{1}{2}}\varphi_{(c)}\varphi_{(c)}^{,i}\,_{,i}
-\frac{3\lambda(\lambda+1)P_{T}}{8a^{3}V}\int
d^{3}x\gamma^{\frac{1}{2}}\varphi_{(c)}^{2} \nonumber \\
&-&\frac{\sqrt{\lambda}}{2}\sqrt{\frac{3}{2}}\frac{l\sqrt{(\lambda+1)P_{T}}}{a^{3}\sqrt{V}}\int
d^{3}x\varphi_{(c)}\pi_{\psi}\,_{(c)}+\frac{a}{3l^{2}}\int
d^{3}x\gamma^{\frac{1}{2}}\psi\psi^{,i}\,_{,i}
\end{eqnarray}
and $\phi_{6}$ by (we omitted the bars)
\begin{equation}
\label{fi6-canonico-hidro}
\phi_{6}=\frac{\sqrt{6}l\sqrt{(\lambda+1)P_{T}}}{\sqrt{\lambda}\sqrt{V}}a^{-3\lambda}\pi_{\varphi}\,_{(c)}-
\frac{l^{2}P_{a}}{2a^{2}V}\pi_{\psi}\,_{(c)}
-\frac{3\sqrt{\lambda}lP_{a}\sqrt{(\lambda+1)P_{T}}}{2\sqrt{6}a^{2}V^{\frac{3}{2}}}\gamma^{\frac{1}{2}}\varphi_{(c)}
+ \frac{2a}{3l^{2}}\gamma^{\frac{1}{2}}\psi^{,i}\,_{,i}
\end{equation}

From the second class constraints we obtain the identity,
\begin{equation}
\label{vinculo-segunda-classe-hidro}
\frac{l^{2}}{2a^{2}\gamma^{\frac{1}{2}}}\pi_{\psi}+\frac{1}{3}F^{,i}\,_{,i}=\frac{3l^{2}}{2}
\frac{\sqrt{\lambda}\sqrt{(\lambda+1)P_{T}}}{\sqrt{6}l\sqrt{V}}a^{-\frac{3}{2}(1+\lambda)}\varphi,
\end{equation}
which in terms of the new canonical variables reads
\begin{equation}
\label{vinculo-segunda-classe-canonico-hidro}
\frac{l^{2}}{2a^{2}\gamma^{\frac{1}{2}}}\pi_{\psi}\,_{(c)}+\frac{1}{3}F^{,i}\,_{,i}=0.
\end{equation}
We will need this equation later.

If we now perform a canonical transformation generated by
\begin{equation}
\label{gerador2-hidro} F_{1}=T{\tilde{P}}_T+a\tilde{P_{a}}+\int
d^{3}x\biggl[\frac{1}{\sqrt{6}l}a^{-\frac{1}{2}(1-3\lambda)}{\tilde{\varphi}}_{(c)}\pi+
\psi{\tilde{\pi}}_{\psi}
+\frac{2\sqrt{V}\sqrt{(\lambda+1){\tilde{P}}_{T}}}{l^{2}\tilde{P_{a}}\sqrt{\lambda}}
a^{\frac{3}{2}(1-\lambda)}\psi\pi-\frac{\gamma^{\frac{1}{2}}}{2}\alpha{\tilde{\varphi}}_{(c)}^{2}\biggr]
\end{equation}
where $\alpha$ is given by
\begin{equation}
\label{alfa-hidro}
\alpha=-\frac{(\lambda+1){\tilde{P}}_{T}}{2l^{2}\tilde{P_{a}}\:a}+\frac{(1-3\lambda)
\tilde{P_{a}}}{24V}a^{-(2-3\lambda)},
\end{equation}
constructed in order to introduce the $v$ variable of
Ref.~\cite{MFB} ($\pi$ is its canonical momentum)\footnote{The term
proportional to $\alpha$ in Eq.~(\ref{gerador2-hidro}), as well as
the specific form given by Eq.~(\ref{alfa-hidro}) are made in order
to eliminate a term proportional to $v\pi$ in the hamiltonian.}, the
new $H_{0}$ reads (the explicit canonical transformations are given
in Appendix B)
\begin{eqnarray}
\label{h0-afa-hidro}
H_{0}&=&-\frac{l^{2}P_{a}^{2}}{4aV}+\frac{P_{T}}{a^{3\lambda}}+\frac{1}{2a}\int
d^{3}x\frac{\pi^{2}}{\gamma^{\frac{1}{2}}}+ \frac{\lambda}{2a}\int
d^{3}x\gamma^{\frac{1}{2}}v^{,i}v_{,i}
+\biggl[\frac{9[(\lambda+1)P_{T}]^{2}}{2P_{a}^{2}}a^{-(1+6\lambda)}
-\frac{9\lambda[(\lambda+1)P_{T}]P_{T}}{2P_{a}^{2}}a^{-(1+6\lambda)}
\nonumber \\
&+&\frac{(-4+18\lambda-18\lambda^{2})l^{4}P_{a}^{2}}{64a^{3}V^{2}}
+\frac{6l^{2}(1+3\lambda^{2})P_{T}}{16V}a^{-(2+3\lambda)}\biggr]\int
d^{3}x\gamma^{\frac{1}{2}}v^{2}
+\frac{2\sqrt{\lambda}\sqrt{V}\sqrt{(\lambda+1)P_{T}}}{l^{2}P_{a}}a^{\frac{1}{2}(1-3\lambda)}\int
d^{3}x\gamma^{\frac{1}{2}}v\psi^{,i}\,_{,i} \nonumber \\
&-&\frac{\sqrt{\lambda}}{2}\frac{3l^{2}\sqrt{(\lambda+1)P_{T}}}{\sqrt{V}}a^{-\frac{1}{2}(5+3\lambda)}\int
d^{3}xv\pi_{\psi}+\int
d^{3}x\psi\biggl\{\biggl[\frac{18[(\lambda+1)P_{T}]^{3}V}{l^{4}P_{a}^{4}\lambda}a^{(2-9\lambda)}
-\frac{18[(\lambda+1)P_{T}]^{2}P_{T}V}{l^{4}P_{a}^{4}}a^{(2-9\lambda)}
\nonumber \\
&+&\frac{(-2+9\lambda-9\lambda^{2})[(\lambda+1)P_{T}]}{8\lambda
V}a^{-3\lambda}
-\frac{3[(\lambda+1)P_{T}]^{2}}{l^{2}P_{a}^{2}\lambda}a^{(1-6\lambda)}
+\frac{3(3+2\lambda+3\lambda^{2})[(\lambda+1)P_{T}]P_{T}}{2l^{2}P_{a}^{2}\lambda}a^{(1-6\lambda)}\biggr]\gamma^{\frac{1}{2}}\psi
\nonumber \\
&+&\biggl[\frac{(-2+3\lambda)\sqrt{(\lambda+1)P_{T}}}{2\sqrt{\lambda}\sqrt{V}}a^{-\frac{1}{2}(1+3\lambda)}
-\frac{6\lambda\sqrt{V}\sqrt{(\lambda+1)P_{T}}P_{T}}{l^{2}P_{a}^{2}\sqrt{\lambda}}a^{\frac{1}{2}(1-9\lambda)}\biggr]\pi
\nonumber \\
&+&\biggl[-\frac{18[(\lambda+1)P_{T}]^{\frac{5}{2}}\sqrt{V}}{l^{2}P_{a}^{3}\sqrt{\lambda}}a^{\frac{1}{2}(1-15\lambda)}
+\frac{18\lambda[(\lambda+1)P_{T}]^{\frac{3}{2}}P_{T}\sqrt{V}}{l^{2}P_{a}^{3}\sqrt{\lambda}}a^{\frac{1}{2}(1-15\lambda)}
+\frac{(2-9\lambda+9\lambda^{2})l^{2}P_{a}\sqrt{(\lambda+1)P_{T}}}{8V^{\frac{3}{2}}\sqrt{\lambda}}a^{-\frac{3}{2}(1+\lambda)}
\nonumber \\
&+&\frac{3[(\lambda+1)P_{T}]^{\frac{3}{2}}}{\sqrt{\lambda}\sqrt{V}P_{a}}a^{-\frac{1}{2}(1+9\lambda)}
-\frac{3(3+2\lambda+3\lambda^{2})\sqrt{(\lambda+1)P_{T}}P_{T}}{2\sqrt{\lambda}\sqrt{V}P_{a}}a^{-\frac{1}{2}(1+9\lambda)}\biggr]\gamma^{\frac{1}{2}}v
\nonumber \\
&+&\biggl[\frac{a}{3l^{2}}-\frac{2(\lambda+1)P_{T}V}{l^{4}P_{a}^{2}}a^{2-3\lambda}\biggr]\gamma^{\frac{1}{2}}\psi^{,i}\,_{,i}
+\frac{3(\lambda+1)P_{T}}{P_{a}}a^{-(1+3\lambda)}\pi_{\psi}\biggr\}
\end{eqnarray}

This canonical transformation applied to $\phi_{6}$ shows that $v$
is a gauge invariant quantity. The same is not true for its momentum $\pi$, which has a non-zero Poisson Bracket with the first class
constraint $\phi_6$. In order to obtain a gauge invariant momentum $\pi$ we now make the canonical transformation generated by
\begin{eqnarray}
\label{gerador3-hidro}
\mathcal{F}_{2}&=&a\tilde{P_{a}}+\int d^{3}x\{\psi\tilde{\pi_{\psi}}+v\tilde{\pi}
+\biggl[\frac{6\sqrt{V}[(\lambda+1)P_{T}]^{\frac{3}{2}}}{l^{2}\tilde{P_{a}}^{2}\sqrt{\lambda}}a^{\frac{3}{2}(1-3\lambda)}
-\frac{(1+3\lambda)\sqrt{(\lambda+1)P_{T}}}{2\sqrt{\lambda}\sqrt{V}}a^{\frac{1}{2}(1-3\lambda)}\biggr]\gamma^{\frac{1}{2}}v\psi \nonumber\\
&+&\biggl[-\frac{6V[(\lambda+1)P_{T}]^{2}}{l^{4}\tilde{P_{a}}^{3}\lambda}a^{3(1-2\lambda)}
+\frac{(1+3\lambda)(\lambda+1)P_{T}}{2l^{2}\tilde{P_{a}}\lambda}a^{2-3\lambda}\biggr]\gamma^{\frac{1}{2}}\psi^{2}
+\frac{2a^{3}V}{3l^{4}\tilde{P_{a}}}\psi\psi^{,i}\:_{,i}\biggr\},
\end{eqnarray}
\end{widetext}
and aiming at eliminating a term in $v^2$ proportional to $P_T$ in
the final form of the hamiltonian, we perform the last canonical
transformation
\begin{equation}
\label{gerador4-hidro}
\mathcal{F}_{3}=a\tilde{P}_{a}+\frac{1}{a}\int d^{3}x
v\tilde{\pi}-\frac{l^{2}\tilde{P}_{a}}{4aV}\int
d^{3}x\gamma^{\frac{1}{2}}v^{2}.
\end{equation}

The constraint $\phi_{6}$ reads then
\begin{equation}
\label{fi6-transf3-hidro}
\phi_{6}=-\frac{l^{2}P_{a}}{2a^{2}V}\pi_{\psi},
\end{equation}
and as $-\frac{l^{2}P_{a}}{2a^{2}V}$ is not weakly zero, we can
redefine the constraint $\phi_6$ to be
\begin{equation}
\phi_6=\pi_\psi
\end{equation}
This constraint, in the Dirac quantization scheme, will imply that
the wave functional will not depend on $\psi$, and the second class
constraint (\ref{vinculo-segunda-classe-canonico-hidro}) turns out
to be the usual relation between $\phi$ and $v$ from
Ref.~\cite{MFB}, Eq.~(12.8), as we will see later on. The new
$H_{0}$ is given by
\begin{widetext}
\begin{eqnarray}
\label{h0-transf4-hidro}
H_{0}&=&-\frac{l^{2}P_{a}^{2}}{4aV}+\frac{P_{T}}{a^{3\lambda}}+\frac{1}{2a^{3}}\int
d^{3}x\frac{\pi^{2}}{\gamma^{\frac{1}{2}}}+\frac{a\lambda}{2} \int
d^{3}x\gamma^{\frac{1}{2}}v^{,i}v_{,i}\nonumber\\
&+&\biggl\{-\frac{\sqrt{\lambda}}{2}\frac{3l^{2}\sqrt{(\lambda+1)P_{T}}}{\sqrt{V}}a^{-\frac{3}{2}(1+\lambda)}v
+\frac{3(\lambda+1)P_{T}}{P_{a}}a^{-(1+3\lambda)}\psi\biggr\}\int
d^{3}x\pi_{\psi} \nonumber \\ &+&H_{0}^{(0)}\int
d^{3}x\biggl\{-\frac{9(\lambda+1)P_{T}}{l^{2}P_{a}^{2}}a^{1-3\lambda}\gamma^{\frac{1}{2}}\psi^{2}
+\frac{6\sqrt{V}\sqrt{(\lambda+1)P_{T}}}{l^{2}P_{a}^{2}\sqrt{\lambda}}a^{-\frac{1}{2}(1+3\lambda)}\psi\pi
\nonumber \\
&+&\biggl[\frac{18[(\lambda+1)P_{T}]^{\frac{3}{2}}\sqrt{V}}{l^{2}P_{a}^{3}\sqrt{\lambda}}a^{\frac{3}{2}(1-3\lambda)}
+\frac{9(\lambda-1)\sqrt{(\lambda+1)P_{T}}}{2\sqrt{V}\sqrt{\lambda}P_{a}}a^{\frac{1}{2}(1-3\lambda)}\biggr]\gamma^{\frac{1}{2}}v\psi
\nonumber\\&+&\frac{2a^{2}V}{l^{4}P_{a}^{2}}\psi\psi^{,i}\,_{,i}
+\biggl[-\frac{9(1+2\lambda)[(\lambda+1)P_{T}}{2P_{a}^{2}}a^{1-3\lambda}
+\frac{9\lambda(\lambda-1)l^{2}}{8V}\biggr]v^{2}\biggr\}
\nonumber\\&\equiv& H_0^{(0)} + H_0^{(2)} + \int d^{3}x
\biggl[F^{(1)}\phi_6 +F^{(2)}H_0^{(0)}\biggr],
\end{eqnarray}
\end{widetext}
where $H_0^{(0)}$ and  $H_0^{(2)}$ are the zeroth and second order
hamiltonian constraints, and $F^{(1)}$ and $F^{(2)}$ are first and
second order functions which can be read from
Eq.~(\ref{h0-transf4-hidro}).

We now make the redefinitions of $N$ and $\phi$ as
$\tilde{N}=N(1+\int d^3x F^{(2)})$, which would again just imply a
different irrelevant time gauge choice with terms beyond first
order, and $\tilde\phi=N(-\frac{l^2P_a}{2a^2V}\phi +F^{(1)})$.
From the inverse of the transformations (\ref{gerador4-hidro}) and
(\ref{canonicas-escalares-hidro}), and definition
(\ref{def-fi-escalares-hidro}), $\tilde{\phi}$ is given by
\begin{equation}
\frac{l^2P_a}{2a^2V}\phi-\frac{3l^2}{2}Na(\lambda+1)\epsilon_0(\frac{a}{N}\dot{\xi}+B)
\end{equation}
As $\tilde{\phi}$ is, through the equations of motion, equal to
$\dot{\psi}$, we obtain, imposing $N=a$, the constraint equation
(10.39) of reference \cite{MFB}.

Inserting expression (\ref{h0-transf4-hidro}) into
Eq.~(\ref{h-vinculos-escalares-hidro}), and the above redefinitions,
we obtain, omitting the tilda,
\begin{equation}
\label{hfinal-vinculos-escalares-hidro} H=N(H_0^{(0)} +
H_0^{(2)})+\Lambda_N P_N + \int d^{3}x\phi\phi_{6}+ \int
d^{3}x\Lambda_{\phi}\pi_{\phi},
\end{equation}
with
\begin{equation}
\label{h00} H_0^{(0)}\equiv
-\frac{l^{2}P_{a}^{2}}{4aV}+\frac{P_{T}}{a^{3\lambda}},
\end{equation}
and
\begin{equation}
\label{h02} H_0^{(2)}\equiv \frac{1}{2a^{3}}\int
d^{3}x\frac{\pi^{2}}{\gamma^{\frac{1}{2}}}+\frac{a\lambda}{2} \int
d^{3}x\gamma^{\frac{1}{2}}v^{,i}v_{,i}.
\end{equation}

Now we are left with two first class constraints (in fact one plus
$\infty^3$ constraints): one with the homogeneous lapse function
$N$ as its associated Lagrange multiplier, which in the
quantization procedure will lead to the Wheeler-DeWitt equation,
and the other $\infty^3$ constraints with $\phi(x^i)$ as their
Lagrange multiplier, which is nothing but the inhomogeneous lapse
function (see definition (\ref{perturb-componentes})), which, as
anticipated, has been tremendously simplified to imply a simple
consequence when quantized: the wave functional does not depend on
$\psi$. The super-momentum constraint is automatically satisfied
because the $v$ variable is gauge invariant.

The connection between $\bar{v}=av$ (the Mukhanov-Sasaki variable)
and $\Phi$ can be obtained from Eq.~
(\ref{vinculo-segunda-classe-canonico-hidro}) which, after
implementing the canonical transformations (\ref{gerador2-hidro},
\ref{gerador3-hidro}) reads (the bars are omitted)
\begin{widetext}
\begin{eqnarray}
\label{vicnulo-transformado}
&&\frac{l^2}{2a^2\gamma^{\frac{1}{2}}}\pi_\psi+\frac{2aV}{3l^2P_a}\biggl(\psi+\frac{l^2P_a}{2aV}F\biggr)^{,i}\,_{,i}
+\biggl[\frac{3\sqrt{V}[(\lambda+1)P_T]^\frac{3}{2}}{P_a^2\sqrt{\lambda}}a^{-\frac{1}{2}(1+9\lambda)}
-\frac{(1+3\lambda)\sqrt{(\lambda+1)P_T}l^2}{4\sqrt{\lambda}\sqrt{V}}a^{-3/2(1+\lambda)}\biggr]v
\nonumber\\&&+\frac{\sqrt{(\lambda+1)P_T}\sqrt{V}}{P_a\sqrt{\lambda}\gamma^{\frac{1}{2}}}a^{-\frac{1}{2}(1+3\lambda)}\pi=0
\end{eqnarray}

Using that $P_a=-\frac{2Va\dot{a}}{Nl^2}$, we can identify the
quantity $\psi+\frac{l^2P_a}{2aV}F$ with the Bardeen potential
$\Psi$ which, from Eq.~(\ref{fi-psi-hidro}), is equal to $\Phi$.
After some algebraic manipulations, we obtain:
\begin{equation}
\label{vicnulo-transformado2}
\frac{3l^4P_a}{4a^3V\gamma^{\frac{1}{2}}}\pi_\psi+\Phi^i\,_i
+\frac{3l^2\sqrt{(\lambda+1)P_T}}{2\sqrt{\lambda}\sqrt{V}}a^{-\frac{3}{2}(1+\lambda)}
\biggl\{\frac{\pi}{\gamma^{\frac{1}{2}}}+\biggl[(\frac{l^2P_a}{2aV}+\frac{3(\lambda+1)}{P_a}H_0)v\biggr]\biggr\}
+O(3)=0
\end{equation}
\end{widetext}
Using again that $P_a=-\frac{2Va\dot{a}}{Nl^2}$, and that
$\pi=\gamma^{\frac{1}{2}}\dot{v}$, $H_0\approx 0$,
$\pi_\psi\approx0$, and choosing the gauge $N=a$ (conformal time),
we obtain
\begin{eqnarray}
\label{vicnulo-simples}
&&\Phi^{,i}\,_{,i}=-\frac{3l^2\sqrt{(\lambda+1)P_T}}{2\sqrt{\lambda}\sqrt{V}}a^{-\frac{1}{2}(1+3\lambda)}\biggl(\frac{v}{a}\biggr)'\nonumber \\
\end{eqnarray}

Equation (\ref{vicnulo-simples}) coincides with equation (12.8) of
Ref.~\cite{MFB} relating $v$ and $\Phi$ when the classical equations
of motion are used.

\section{Dirac quantization}

In this section we will focus only in the quantization of scalar
perturbations. Vector perturbations are trivial and the quantization
of tensor perturbations was done in Ref.~\cite{tens1}.

\subsection{The functional Schr\"odinger equation}

In the Dirac quantization procedure, the first class constraints
must annihilate the wave functional
$\chi[N,a,\phi(x^i),\psi(x^ï),v(x^i),T]$, yielding
\begin{eqnarray}
\label{vinculos-quanticos}
\frac{\partial}{\partial N}\chi&=&0, \nonumber \\
\frac{\delta}{\delta\phi}\chi&=&0, \nonumber \\
\frac{\delta}{\delta\psi}\chi&=&0, \nonumber \\
H\chi&=&0.
\end{eqnarray}

The first three equations impose that the wave functional does not
depend on $N$, $\phi$ and $\psi$: as mentioned above, $N$ and $\phi$
are, respectively, the homogeneous and inhomogeneous parts of the
total lapse function, which are just lagrange multipliers of
constraints, and $\psi$ has been substituted by $v(x^i)$, the unique
degree of freedom of scalar perturbations, as expected.

As $P_T$ appears linearly in $H$, and making the gauge choice
$N=a^{3\lambda}$, one can interpret the $T$ variable as a time
parameter. Hence, the equation
\begin{equation}
\label{schroedinger} H\chi=0
\end{equation}
assumes the Schr\"odinger form
\begin{widetext}
\begin{equation}
\imath\frac{\partial}{\partial T}\chi =\frac{1}{4} \left\{
a^{(3\lambda-1)/2}\frac{\partial}{\partial a} \left[
a^{(3\lambda-1)/2}\frac{\partial}{\partial a}\right]
\right\}\chi-\biggl[\frac{a^{3\lambda-1}}{2}\int
d^3x\frac{\delta^2}{\delta v^2}-\frac{a^{3\lambda+1}\lambda}{2}\int
d^3x v^{,i}v_{,i}\biggr]\chi ,
\end{equation}
\end{widetext}
where we have chosen the factor ordering in $a$ in order to yield
a covariant Schr\"odinger equation under field redefinitions, and
$V$ and $l$ have been absorbed in redefinitions of the fields.

\subsection{Further developments using the Bohm-de Broglie
interpretation}

If one makes the ansatz
\begin{equation}
\label{ansatz} \chi[a,v,T]=\chi_{(0)}(a,T)\chi_{(2)}[v,T]
\end{equation}
where $\chi_{(0)}(a,T)$ satisfies the equation,
\begin{eqnarray}
\label{scrhoedinger-separado-fundo} &&\imath\frac{\partial}{\partial
T} \chi_{(0)}(a,T)=\nonumber\\&&\frac{1}{4} \left\{
a^{(3\lambda-1)/2}\frac{\partial}{\partial a} \left[
a^{(3\lambda-1)/2}\frac{\partial}{\partial a}\right] \right\}
\chi_{(0)}(a,T),
\end{eqnarray}
then we obtain for $\chi_{(2)}(a,v,T)$ the equation
\begin{widetext}
\begin{equation}
\label{scrhoedinger-separado-perturb} \imath\frac{\partial}{\partial
T} \chi_{(2)}(a,v,T)=-\frac{a^{(3\lambda-1)}}{2}\int
d^3x\frac{\delta^2}{\delta v^2}\chi_{(2)}(a,v,T)+\frac{\lambda
a^{(3\lambda+1)}}{2}\int d^3x v^{,i}v_{,i}\chi_{(2)}(a,v,T)
\end{equation}
\end{widetext}

Solutions of the zeroth order equation
(\ref{scrhoedinger-separado-fundo}) are known \cite{pinto,fabris}.
If one uses the ontological Bohm-de Broglie interpretation of
quantum mechanics in order to obtain the bohmian trajectories $a(T)$
from Eq.~(\ref{scrhoedinger-separado-fundo}), this $a(T)$ can be
viewed as a given function of time in the second equation
(\ref{scrhoedinger-separado-fundo}). Going to conformal time
$d\eta=a^{3\lambda-1}dT$, and performing the unitary transformation
\begin{equation}
\label{unitarias} U=e^{\{\imath[\int
d^3x\gamma^{\frac{1}{2}}\frac{\dot{a}v}{2a}]\}}e^{\{\imath[\int
d^3x(\frac{v\pi+\pi v}{2})\ln (\frac{1}{a})]\}},
\end{equation}
the Schr\"odinger functional equation for the perturbations is
transformed to
\begin{equation}
i\frac{\partial\chi_{(2)}[v,\eta]}{\partial \eta}= \int \dd^3 x
\left(-\frac{1}{2} \frac{\delta^2}{\delta v^2} +
\frac{\lambda}{2}v_{,i} v^{,i} - \frac{{a''}}{2a}v^2 \right)
\chi_{(2)}[v,\eta], \label{schroedinger-conforme}
\end{equation}
where we have gone to the new quantum variable $\bar{v}=av$, the
Mukhanov-Sasaki variable defined in Ref.~\cite{MFB}, after
performing transformation (\ref{unitarias}), and we have omitted the
bars.

The corresponding time evolution equation for the operator $v$ in
the Heisenberg picture is given by
\begin{equation}
\label{equacoes-mukhanov} v''-\lambda
v^{,i}\,_{,i}-\frac{a''}{a}v=0,
\end{equation}
where a prime means derivative with respect to conformal time. In
terms of the normal modes $v_k$, the above equation reads
\begin{equation}
\label{equacoes-mukhanov}
v''_k+\biggl(k^2-\frac{{a''}}{a}\biggr)v_k=0.
\end{equation}

These equations have the same form as the equations for scalar
perturbations obtained in Ref.~\cite{MFB} (for one single fluid, the
pump function $z''/z$ obtained in \cite{MFB} is exactly equal to
$a''/a$ obtained here, if we make use of the background equations).
The difference is that the function $a(\eta)$ is no longer a
classical solution of the background equations but a quantum Bohmian
trajectory of the quantized background, which may lead to different
power spectra.

\section{Conclusion}

In this paper we have managed to obtain simple hamiltonians for
scalar perturbations when the matter content is described either
by a perfect fluid or by a scalar fluid, without recurring to the
background classical equations. Performing canonical
transformations and redefining the homogeneous lapse functions
with terms which do not alter the linear perturbation equations,
the constraint connected to the inhomogeneous part of the lapse
function is greatly simplified implying that the momentum
canonically conjugate to the scalar perturbation $\psi$ is weakly
zero. The hamiltonian constraint is also greatly simplified when
written in terms of a new variable which is exactly equal to the
usual Mukhanov-Sasaki's variable \cite{MFB}.

This simplified hamiltonian can now be used in the Dirac
quantization procedure not only to quantize the perturbations but
also the background, yielding a Wheeler-DeWitt equation much
simpler to handle then the one of Ref.~\cite{halli}. In the case
of perfect fluids, where a preferred time variable appears and the
Wheeler-DeWitt equation can be put in a Schr\"odinger form, and
using the Bohm-de Broglie interpretation of quantum mechanics to
perform a last unitary transformation, one obtains an equation for
the modes which has the same form as in Ref.~\cite{MFB}, where the
pump field is obtained from a scale factor which now takes into
account the quantum effects, the quantum Bohmian trajectory of the
background.

In future publications, we will apply these results to specific
models, and evaluate the power spectrum of scalar perturbations
which arise on them in order to compare, when taken together with
the results of Ref.~\cite{tens2}, with observations.

\section{Acknowledgments}

We would like to thank CNPq of Brazil, CAPES (Brazil) and COFECUB
(France) for financial support. We very gratefully acknowledge
various enlightening conversations with Patrick Peter.
\newpage
\appendix

\section{The scalar field}

In this appendix we implement the same simplifications we have done
for hydrodynamical matter for the case of a scalar field.

The scalar field lagrangian reads
\begin{equation}
\label{lagrange-densidade-escalar}
\pounds_{m}=\frac{1}{2}\varphi_{;\mu}\varphi^{;\mu}-\frac{1}{2}U(\varphi).
\end{equation}
We write its perturbation as
\begin{equation}
\label{perturb-escalar} \varphi=\varphi_{0}+\delta\varphi,
\end{equation}
where $\varphi_{0}$ is the homogeneous scalar field depending only
on time. Substituting \ref{perturb-escalar} in
\ref{lagrange-densidade-escalar} we obtain
\begin{widetext}
\begin{eqnarray}
\label{lagrange-perturb-densidade-escalar}
&\pounds_{m}=&\frac{\dot{\varphi_{0}}^{2}}{2N^{2}}-\frac{1}{2}U-\frac{\dot{\varphi_{0}}^{2}}{N^{2}}\phi+\frac{\dot{\varphi_{0}}}{N^{2}}\dot{\delta\varphi}-\frac{1}{2}U_{\varphi}\delta\varphi+\frac{2\dot{\varphi_{0}}^{2}}{N^{2}}\phi^{2}-\frac{\dot{\varphi_{0}}^{2}}{2N^{2}}A^{i}A_{i}-\frac{2\dot{\varphi_{0}}}{N^{2}}\phi\dot{\delta\varphi}-\frac{\dot{\varphi_{0}}}{Na}A^{i}\delta\varphi_{|i}
\nonumber \\
&&\frac{1}{2N^{2}}\dot{\delta\varphi}^{2}-\frac{1}{2a^{2}}\delta\varphi^{i}\delta\varphi_{i}-\frac{1}{4}U_{\varphi\varphi}\delta\varphi^{2}
\end{eqnarray}

The total lagrangian including the gravitational sector reads
\begin{eqnarray}
\label{lagrange-total-esc}
L&=&-\frac{\dot{a}^{2}aV}{l^{2}N}+\frac{NKaV}{l^{2}}+\frac{\dot{\varphi_{0}}^{2}a^{3}V}{2N}-\frac{Na^{3}VU}{2}+\frac{Na}{6l^{2}}\int d^{3}x\gamma^{\frac{1}{2}}[A^{i|j}A_{[i|j]}-\frac{1}{4}\epsilon^{ij|k}\epsilon_{ij|k}\nonumber \\
&&+\frac{a}{N}\dot{A_{i}}\epsilon^{ij}\,_{|j}+\frac{1}{2}\epsilon^{ij}\,_{|j}\epsilon_{i}\,^{k}\,_{|k}+\phi_{|i}\epsilon^{ij}\,_{|j}-\frac{1}{2}\epsilon_{|i}\epsilon^{ij}\,_{|j}-\phi_{|i}\epsilon^{|i}+\frac{1}{4}\epsilon_{|i}\epsilon^{|i}+K(\frac{1}{4}\epsilon^{2}-\epsilon^{ij}\epsilon_{ij}-\epsilon\phi \nonumber \\
&&+A^{i}A_{i}-3\phi^{2})]+\frac{a^{3}}{24l^{2}N}\int
d^{3}x\gamma^{\frac{1}{2}}\dot{\epsilon^{ij}}\dot{\epsilon_{ij}}-\frac{a^{3}}{24l^{2}N}\int d^{3}x\gamma^{\frac{1}{2}}\dot{\epsilon}^{2}+\frac{a\dot{a}^{2}}{6l^{2}N}\int d^{3}x\gamma^{\frac{1}{2}}(-9\phi^{2}-3\epsilon\phi \nonumber \\
&&-\frac{3}{4}\epsilon^{2}+3A^{i}A_{i}+\frac{3}{2}\epsilon^{ij}\epsilon_{ij})-\frac{2a\dot{a}}{3l^{2}}\int
d^{3}x\gamma^{\frac{1}{2}}(\phi A^{i}\,_{|i}-\frac{1}{2}A_{i}\epsilon^{ij}\,_{|j})+\frac{a^{2}\dot{a}}{3l^{2}N}\int d^{3}x\gamma^{\frac{1}{2}}(\epsilon^{ij}\dot{\epsilon_{ij}}-\frac{1}{2}\epsilon\dot{\epsilon}-\phi\dot{\epsilon}) \nonumber \\
&&-\frac{a^{2}}{6l^{2}}\int
d^{3}x\gamma^{\frac{1}{2}}\dot{\epsilon}A^{i}\,_{|i}-\frac{a^{3}\dot{\varphi_{0}}}{N}\int
d^{3}x\gamma^{\frac{1}{2}}(\phi+\frac{1}{2}\epsilon)\dot{\delta\varphi}+a^{2}\dot{\varphi_{0}}\int d^{3}x\gamma^{\frac{1}{2}}\delta\varphi A^{i}\,_{|i}-\frac{Na^{3}U_{\varphi}}{2}\int d^{3}x\gamma^{\frac{1}{2}}(\phi-\frac{1}{2}\epsilon)\delta\varphi  \nonumber \\
&&+\frac{\dot{\varphi_{0}}^{2}a^{3}}{4N}\int
d^{3}x\gamma^{\frac{1}{2}}(3\phi^{2}+\epsilon\phi-A^{i}A_{i}-\frac{1}{2}\epsilon^{ij}\epsilon_{ij}+\frac{1}{4}\epsilon^{2})+\frac{Na^{3}U}{4}\int d^{3}x\gamma^{\frac{1}{2}}(\phi^{2}+\epsilon\phi-A^{i}A_{i}+\frac{1}{2}\epsilon^{ij}\epsilon_{ij}-\frac{1}{4}\epsilon^{2}) \nonumber \\
&&+\frac{Na^{3}}{2}\int
d^{3}x\gamma^{\frac{1}{2}}(\frac{\dot{\delta\varphi}^{2}}{N^{2}}-\frac{\delta\varphi^{|i}\delta\varphi_{|i}}{a^{2}}-\frac{1}{2}U_{\varphi\varphi}\delta\varphi^{2})
\end{eqnarray}
Its hamiltonian is given by
\begin{eqnarray}
\label{h-total-esc}
H&=&N\biggr\{-\frac{l^{2}P_{a}^{2}}{4aV}-\frac{KaV}{l^{2}}+\frac{P_{\varphi}^{2}}{2a^{3}V}+\frac{a^{3}VU}{2}+\frac{l^{2}P_{a}^{2}}{8aV^{2}}\int
d^{3}x\gamma^{\frac{1}{2}}\phi^{2}+\frac{l^{2}P_{a}^{2}}{24aV^{2}}\int d^{3}x\gamma^{\frac{1}{2}}\epsilon\phi-\frac{l^{2}P_{a}^{2}}{8aV^{2}}\int d^{3}x\gamma^{\frac{1}{2}}A^{i}A_{i} \nonumber \\
&&+\frac{5l^{2}P_{a}^{2}}{48aV^{2}}\int
d^{3}x\gamma^{\frac{1}{2}}\epsilon^{ij}\epsilon_{ij}-\frac{l^{2}P_{a}^{2}}{32aV^{2}}\int
d^{3}x\gamma^{\frac{1}{2}}\epsilon^{2}+\frac{P_{a}}{6V}\int d^{3}x\gamma^{\frac{1}{2}}A^{i}\,_{|i}\phi+\frac{P_{a}}{6V}\int d^{3}x\gamma^{\frac{1}{2}}A_{i}\epsilon^{ij}\,_{|j} \nonumber \\
&&-\frac{P_{\varphi}^{2}}{4a^{3}V^{2}}\int
d^{3}x\gamma^{\frac{1}{2}}(\phi^{2}-\epsilon\phi-A^{i}A_{i}-\frac{1}{2}\epsilon^{ij}\epsilon_{ij}-\frac{1}{4}\epsilon^{2})+\frac{P_{\varphi}}{a^{3}V}\int
d^{3}x(\phi+\frac{1}{2}\epsilon)\pi_{\varphi}-\frac{P_{\varphi}}{aV}\int d^{3}x\gamma^\frac{1}{2}\delta\varphi A^{i}\,_{|i} \nonumber \\
&&+\frac{6l^{2}}{a^{3}}\int
d^{3}x\frac{\pi^{ij}\pi_{ij}}{\gamma^{\frac{1}{2}}}-\frac{3l^{2}}{a^{3}}\int
d^{3}x\frac{\pi^{2}}{\gamma^{\frac{1}{2}}}-\frac{l^{2}P_{a}}{2a^{2}V}\int
d^{3}x \pi\epsilon-\frac{a}{4l^{2}}\int
d^{3}x\gamma^{\frac{1}{2}}A^{i}\,_{|i}A^{j}\,_{|j}-\frac{1}{a}\int d^{3}x \pi A^{i}\,_{|i} \nonumber \\
&&+\frac{l^{2}P_{a}}{a^{2}V}\int
d^{3}x\pi\phi+\frac{2l^{2}P_{a}}{a^{2}V}\int
d^{3}x\pi^{ij}\epsilon_{ij}+\frac{1}{2a^{3}}\int d^{3}x\frac{\pi_{\varphi}^{2}}{\gamma^{\frac{1}{2}}}-\frac{a}{6l^{2}}\int d^{3}x\gamma^{\frac{1}{2}}\biggr[A^{i|j}A_{[i|j]}-\frac{1}{4}\epsilon^{ij|k}\epsilon_{ij|k}+\frac{1}{2}\epsilon^{ij}\,_{|j}\epsilon_{i}\,^{k}\,_{|k} \nonumber \\
&&+\phi_{|i}\epsilon^{ij}\,_{|j}-\frac{1}{2}\epsilon_{|i}\epsilon^{ij}\,_{|j}-\phi_{|i}\epsilon^{|i}+\frac{1}{4}\epsilon^{|i}\epsilon_{|i}+K(\frac{1}{4}\epsilon^{2}-\epsilon^{ij}\epsilon_{ij}-\epsilon\phi+A^{i}A_{i}-3\phi^{2})\biggl]+\frac{a^{3}U_{\varphi}}{2}\int d^{3}x \gamma^{\frac{1}{2}}(\phi-\frac{1}{2}\epsilon)\delta\varphi \nonumber \\
&&-\frac{a^{3}U}{4}\int
d^{3}x\gamma^{\frac{1}{2}}(\phi^{2}+\epsilon\phi-A^{i}A_{i}+\frac{1}{2}\epsilon^{ij}\epsilon_{ij}-\frac{1}{4}\epsilon^{2})+\frac{a^{3}}{2}\int
d^{3}x\gamma^{\frac{1}{2}}(\frac{1}{a^{2}}\delta\varphi^{|i}\delta\varphi_{|i}+\frac{1}{2}U_{\varphi\varphi}\delta\varphi^{2})+\frac{P_{a}}{12V}\int d^{3}x\gamma^{\frac{1}{2}}\epsilon A^{i}\,_{i}\biggl\} \nonumber \\
&&
\end{eqnarray}

Performing the canonical transformation generated by
\begin{eqnarray}
\label{gerador1-esc}
F&=&a\tilde{P_{a}}+\varphi_{0}\tilde{P_{\varphi}}-\int
d^{3}x(\tilde{\phi}\pi_{\phi}+\tilde{A_{i}}\pi_{A}^{i}+\tilde{\epsilon_{ij}}\pi^{ij}+
\tilde{\delta\varphi}\pi_{\varphi})-\frac{a\tilde{P_{a}}}{12V}\int
d^{3}x\gamma^{\frac{1}{2}}(\tilde{\epsilon^{ij}}\tilde{\epsilon_{ij}}-\frac{1}{2}\tilde{\epsilon}^{2})-\frac{\tilde{P_{\varphi}}}{V}\int
d^{3}x\gamma^{\frac{1}{2}}(\tilde{\phi}+\frac{1}{2}\tilde{\epsilon})\tilde{\delta\varphi} \nonumber \\
&&
\end{eqnarray}
which are
\begin{eqnarray}
\label{transf1-esc}
a&=&\tilde{a}+\frac{\tilde{a}}{12V}\int d^{3}x\gamma^{\frac{1}{2}}(\tilde{\epsilon^{ij}}\tilde{\epsilon_{ij}}-\frac{1}{2}\tilde{\epsilon}^{2}) \nonumber \\
P_{a}&=&\tilde{P_{a}}-\frac{\tilde{P_{a}}}{12V}\int d^{3}x\gamma^{\frac{1}{2}}(\tilde{\epsilon^{ij}}\tilde{\epsilon_{ij}}-\frac{1}{2}\tilde{\epsilon}^{2}) \nonumber \\
\varphi_{0}&=&\tilde{\varphi_{0}}+\frac{1}{V}\int d^{3}x\gamma^{\frac{1}{2}}(\tilde{\phi}+\frac{1}{2}\tilde{\epsilon})\tilde{\delta\varphi} \nonumber \\
\pi_{\phi}&=&\tilde{\pi_{\phi}}-\frac{\tilde{P_{\varphi}}}{V}\gamma^{\frac{1}{2}}\tilde{\delta\varphi} \nonumber \\
\pi^{ij}&=&\tilde{\pi^{ij}}-\frac{\tilde{a}\tilde{P_{a}}}{6V}\gamma^{\frac{1}{2}}(\tilde{\epsilon^{ij}}-\frac{1}{2}\tilde{\epsilon}\gamma^{ij})-\frac{\tilde{P_{\varphi}}}{2V}\gamma^{\frac{1}{2}}\tilde{\delta{\varphi}}\gamma^{ij} \nonumber \\
\pi_{\varphi}&=&\tilde{\pi_{\varphi}}-\frac{\tilde{P_{\varphi}}}{V}\gamma^{\frac{1}{2}}(\tilde{\phi}+\frac{1}{2}\tilde{\epsilon})
\end{eqnarray}
yields the new hamiltonian
\newline
\begin{eqnarray}
\label{h-transf-esc}
H&=&N\biggl\{-\frac{l^{2}P_{a}^{2}}{4aV}-\frac{KaV}{l^{2}}+\frac{P_{\varphi}^{2}}{2a^{3}V}+\frac{a^{3}VU}{2}+\frac{l^{2}P_{a}^{2}}{8aV^{2}}\int
d^{3}x\gamma^{\frac{1}{2}}\phi^{2}+\frac{l^{2}P_{a}^{2}}{8aV^{2}}\int d^{3}x\gamma^{\frac{1}{2}}\epsilon\phi-\frac{l^{2}P_{a}^{2}}{8aV^{2}}\int d^{3}x\gamma^{\frac{1}{2}}A^{i}A_{i} \nonumber \\
&&+\frac{P_{a}}{6V}\int
d^{3}x\gamma^{\frac{1}{2}}A^{i}\,_{|i}\phi+\frac{P_{a}}{6V}\int
d^{3}x\gamma^{\frac{1}{2}}A_{i}\epsilon^{ij}\,_{|j}-\frac{P_{\varphi}^{2}}{4a^{3}V^{2}}\int
d^{3}x\gamma^{\frac{1}{2}}(3\phi^{2}+\epsilon\phi-A^{i}A_{i})+\frac{l^{2}P_{a}}{a^{2}V}\int d^{3}x\pi\phi \nonumber \\
&&+\frac{3l^{2}P_{a}}{a^{3}V}\int
d^{3}x\pi\delta\varphi+\frac{P_{\varphi}}{2aV}\int
d^{3}x\gamma^\frac{1}{2}\delta\varphi
A^{i}\,_{|i}+\frac{6l^{2}}{a^{3}}\int
d^{3}x\frac{\pi^{ij}\pi_{ij}}{\gamma^{\frac{1}{2}}}-\frac{3l^{2}}{a^{3}}\int
d^{3}x\frac{\pi^{2}}{\gamma^{\frac{1}{2}}}-\frac{a}{4l^{2}}\int d^{3}x\gamma^{\frac{1}{2}}A^{i}\,_{|i}A^{j}\,_{|j} \nonumber \\
&&-\frac{1}{a}\int d^{3}x \pi
A^{i}\,_{|i}-\frac{9l^{2}P_{\varphi}^{2}}{4a^{3}V^{2}}\int
d^{3}\gamma^{\frac{1}{2}}\delta\varphi^{2}-\frac{3l^{2}P_{a}P_{\varphi}}{2a^{2}V^{2}}\int d^{3}x\gamma^{\frac{1}{2}}\phi\delta\varphi+\frac{1}{2a^{3}}\int d^{3}x\frac{\pi_{\varphi}^{2}}{\gamma^{\frac{1}{2}}}-\frac{a}{6l^{2}}\int d^{3}x\gamma^{\frac{1}{2}}[A^{i|j}A_{[i|j]} \nonumber \\
&&-\frac{1}{4}\epsilon^{ij|k}\epsilon_{ij|k}+\frac{1}{2}\epsilon^{ij}\,_{|j}\epsilon_{i}\,^{k}\,_{|k}+\phi_{|i}\epsilon^{ij}\,_{|j}-\frac{1}{2}\epsilon_{|i}\epsilon^{ij}\,_{|j}-\phi_{|i}\epsilon^{|i}+\frac{1}{4}\epsilon^{|i}\epsilon_{|i}+K(-\frac{1}{2}\epsilon^{ij}\epsilon_{ij}-\epsilon\phi+A^{i}A_{i}-3\phi^{2})] \nonumber \\
&&+a^{3}U_{\varphi}\int
d^{3}x\gamma^{\frac{1}{2}}\phi\delta\varphi-\frac{a^{3}U}{4}\int
d^{3}x\gamma^{\frac{1}{2}}(\phi^{2}+\epsilon\phi-A^{i}A_{i})+\frac{a^{3}}{2}\int
d^{3}x\gamma^{\frac{1}{2}}(\frac{1}{a^{2}}\delta\varphi^{|i}\delta\varphi_{|i}+\frac{1}{2}U_{\varphi\varphi}\delta\varphi^{2})\biggr\}
\end{eqnarray}
Going back to its corresponding lagrangian, and redefining $N$ as
\begin{equation}
\label{N-redef-esc} N=\tilde{N}\biggr[1+\frac{1}{2V}\int
d^{3}x\gamma^{\frac{1}{2}}(\epsilon\phi+\phi^{2}-A^{i}A_{i})\biggl]
\end{equation}
we obtain
\begin{eqnarray}
\label{lagrange-redef-esc}
L&=&-\frac{\dot{a}^{2}aV}{l^{2}N}+\frac{NKaV}{l^{2}}+\frac{\dot{\varphi_{0}}^{2}a^{3}V}{2N}-\frac{Na^{3}VU}{2}+\frac{Na}{6l^{2}}\int d^{3}x\gamma^{\frac{1}{2}}\biggl[A^{i|j}A_{[i|j]}-\frac{1}{4}\epsilon^{ij|k}\epsilon_{ij|k}+\frac{a}{N}\dot{A_{i}}\epsilon^{ij}\,_{|j}+\frac{1}{2}\epsilon^{ij}\,_{|j}\epsilon_{i}\,^{k}\,_{|k}\nonumber \\
&&+\phi_{|i}\epsilon^{ij}\,_{|j}-\frac{1}{2}\epsilon_{|i}\epsilon^{ij}\,_{|j}-\phi_{|i}\epsilon^{|i}+\frac{1}{4}\epsilon_{|i}\epsilon^{|i}+K(-\frac{1}{2}\epsilon^{ij}\epsilon_{ij}+2\epsilon\phi-2A^{i}A_{i})\biggr]+\frac{a^{3}}{24l^{2}N}\int
d^{3}x\gamma^{\frac{1}{2}}\dot{\epsilon^{ij}}\dot{\epsilon_{ij}}-\frac{a^{3}}{24l^{2}N}\int d^{3}x\gamma^{\frac{1}{2}}\dot{\epsilon}^{2}\nonumber \\
&&-\frac{a\dot{a}^{2}}{l^{2}N}\int
d^{3}x\gamma^{\frac{1}{2}}\phi^{2}-\frac{2a\dot{a}}{3l^{2}}\int
d^{3}x\gamma^{\frac{1}{2}}(\phi A^{i}\,_{|i}-\frac{1}{2}A_{i}\epsilon^{ij}\,_{|j})-\frac{a^{2}\dot{a}}{3l^{2}N}\int d^{3}x\gamma^{\frac{1}{2}}\phi\dot{\epsilon}-\frac{a^{2}}{6l^{2}}\int d^{3}x\gamma^{\frac{1}{2}}\dot{\epsilon}A^{i}\,_{|i}\nonumber \\
&&+\frac{a^{3}\dot{\varphi_{0}}}{N}\int
d^{3}x\gamma^{\frac{1}{2}}(\dot{\phi}+\frac{1}{2}\dot{\epsilon})\delta\varphi+a^{2}\dot{\varphi_{0}}\int
d^{3}x\gamma^{\frac{1}{2}}\delta\varphi
A^{i}\,_{|i}-Na^{3}U_{\varphi}\int d^{3}x\gamma^{\frac{1}{2}}\phi\delta\varphi+\frac{\dot{\varphi_{0}}^{2}a^{3}}{2N}\int d^{3}x\gamma^{\frac{1}{2}}\phi^{2}\nonumber \\
&&+\frac{Na^{3}}{2}\int
d^{3}x\gamma^{\frac{1}{2}}(\frac{\dot{\delta\varphi}^{2}}{N^{2}}-\frac{\delta\varphi^{|i}\delta\varphi_{|i}}{a^{2}}-\frac{1}{2}U_{\varphi\varphi}\delta\varphi^{2})
\end{eqnarray}
\end{widetext}

Splitting as before the perturbations into their tensorial, vector
and scalar parts
\begin{eqnarray}
\label{perturb-modos}
A_{i}&=&B_{|i}+S_{i} \nonumber \\
\epsilon_{ij}&=&2\psi\gamma_{ij}-2E_{|i|j}-F_{i|j}-F_{j|i}+w_{ij},
\end{eqnarray}
with
\begin{eqnarray}
\label{perturb-condicoes-esc}
S^{i}\,_{|i}&=&F^{i}\,_{|i}=0 \nonumber \\
w^{ij}\,_{|j}&=&0 \nonumber \\
w^{i}\,_{i}&=&0
\end{eqnarray}
the lagrangian also splits in tensor, vector and scalar parts. The
tensor part was already treated in Ref.~\cite{tens1}.

The vector part reads
\begin{widetext}
\begin{eqnarray}
\label{lagrange-vetor-esc} L^{(V)}&=&\frac{Na}{6l^{2}}\int d^{3}x\gamma^{\frac{1}{2}}S^{i|j}S_{[i|j]}-\frac{a^{2}}{6l^{2}}\int d^{3}x\gamma^{\frac{1}{2}}S_{i|j}\dot{F^{i|j}}+\frac{a^{3}}{12l^{2}N}\int d^{3}x\gamma^{\frac{1}{2}}\dot{F^{i|j}}\dot{F_{i|j}}+\frac{a^{2}K}{3l^{2}}\int d^{3}x\gamma^{\frac{1}{2}}S_{i}\dot{F^{i}} \nonumber \\
&&-\frac{a^{3}K}{6l^{2}N}\int
d^{3}x\gamma^{\frac{1}{2}}\dot{F^{i}}\dot{F_{i}}-\frac{NaK}{3l^{2}}\int
d^{3}x\gamma^{\frac{1}{2}}S^{i}S_{i}
\end{eqnarray}
\end{widetext}
Using the gauge invariant quantity
\begin{equation}
\label{vetor-invariante-esc} V_{i}=S_{i}-\frac{a}{N}\dot{F_{i}},
\end{equation}
this lagrangian simplify to
\begin{equation}
\label{lagrange-invariante-esc} \frac{Na}{6l^{2}}\int
d^{3}x\gamma^{\frac{1}{2}}(V^{[i|j]}V_{[i|j]}-2KV^{i}V_{i})
\end{equation}

Its associated hamiltonian reads
\begin{equation}
\label{h-vetor-esc} H^{(V)}=\frac{Na}{6l^{2}}\int
d^{3}x\gamma^{\frac{1}{2}}V_{i}(\frac{1}{2}V^{i|j}\,_{|j}+KV_{i})+\int
d^{3}x\Lambda_{i}\pi^{i},
\end{equation}
where we have the constraint
\begin{equation}
\label{momenta-vetor-esc} \pi_{V}^{i}\approx0
\end{equation}

Conservation of the constraint $\pi_{V}^{i}\approx 0$ leads to the
secondary constraint
\begin{equation}
\label{vinculo-vetor-esc} \frac{1}{2}V^{i|j}\,_{|j}+KV_{i}\approx0
\end{equation}
whose conservation fixes the Lagrange multiplier $\Lambda_{V}^{i}$,
which means that both constraints are second class. Defining the
associated Dirac brackets, they become strong equalities, yielding
the well known result for a universe filled only with a scalar
field:
\begin{equation}
\label{perturb-vetor-esc} V^{i}=0.
\end{equation}

In the scalar sector we have
\begin{widetext}
\begin{eqnarray}
\label{lagrange-escalares-esc} L^{(E)}&=&\frac{Na}{6l^{2}}\int
d^{3}x\gamma^{\frac{1}{2}}\biggl[2\psi^{,i}\psi_{,i}-4\phi_{,i}\psi^{,i}+K(-6\psi^{2}+12\phi\psi)\biggr]-\frac{a^{3}}{l^{2}N}\int d^{3}x\gamma^{\frac{1}{2}}\dot{\psi}^{2}-\frac{\dot{a}^{2}a}{l^{2}N}\int d^{3}x\gamma^{\frac{1}{2}}\phi^{2} \nonumber \\
&&-\frac{2a^{2}\dot{a}}{l^{2}N}\int
d^{3}x\gamma^{\frac{1}{2}}\phi\dot{\psi}+\frac{a^{3}\dot{\varphi_{0}}}{N}\int
d^{3}x\gamma^{\frac{1}{2}}(\dot{\phi}+3\dot{\psi})\delta\varphi-Na^{3}U_{\varphi}\int
d^{3}x\gamma^{\frac{1}{2}}\phi\delta\varphi+\frac{\dot{\varphi_{0}}^{2}a^{3}}{2N}\int
d^{3}x\gamma^{\frac{1}{2}}\phi^{2}\nonumber \\
&&+\frac{Na^{3}}{2}\int
d^{3}x\gamma^{\frac{1}{2}}(\frac{1}{N^{2}}\dot{\delta\varphi}^{2}-\frac{1}{a^{2}}\delta\varphi^{,i}\delta\varphi_{,i}-
\frac{1}{2}U_{\varphi\varphi}\delta\varphi^{2})-\frac{2a^{2}}{3l^{2}}\biggl[\dot{\psi}+\frac{\dot{a}}{a}\phi-\frac{3l^{2}\dot{\varphi_{0}}}{2}\delta\varphi \nonumber \\
&&-\frac{1}{2}\frac{N}{a}K(B-\frac{a}{N}\dot{E})\biggr](B-\frac{a}{N}\dot{E})^{,i}\,_{,i}
\end{eqnarray}

When constructing the hamiltonian, we obtain the primary constraints
(here again we define $F=B-a\dot{E}/N$)
\begin{eqnarray}
\label{vinculos-escalares-esc}
\phi_{1}=P_{N}&\approx&0, \nonumber \\
\phi_{2}=\pi_{\phi}-\frac{a^{3}\dot{\varphi_{0}}}{N}\gamma^{\frac{1}{2}}\delta\varphi&\approx&0, \nonumber \\
\phi_{3}=\pi_{F}&\approx&0.
\end{eqnarray}
The hamiltonian reads
\begin{equation}
\label{h-escalares-esc} H=N H_{0}+\Lambda_{N}P_{N}+\int
d^{3}x\Lambda_{F}\pi_{F}+\int
d^{3}x\Lambda_{\phi}(\pi_{\phi}-\frac{P_{\varphi}}{V}\gamma^{\frac{1}{2}}\delta\varphi)
\end{equation}
where $H_{0}$ is given by
\begin{eqnarray}
\label{h0-escalares-esc}
H_{0}&=&-\frac{l^{2}P_{a}^{2}}{4aV}+\frac{P_{\varphi}^{2}}{2a^{3}V}-\frac{KaV}{l^{2}}+\frac{a^{3}VU}{2}+\frac{l^{2}P_{a}}{2a^{2}V}
\int d^{3}\phi\pi_{\psi}-\frac{P_{\varphi}^{2}}{2a^{3}V^{2}}\int
d^{3}x\gamma^{\frac{1}{2}}\phi^{2}+\frac{3l^{2}P_{\varphi}}{2a^{3}V}\int
d^{3}\delta\varphi\pi_{\psi}\nonumber \\
&&-\frac{a^{3}}{l^{2}}\int
d^{3}x\gamma^{\frac{1}{2}}(\frac{l^{2}}{2a^{3}\gamma^{\frac{1}{2}}}\pi_{\psi}+\frac{1}{3a}F^{,i}\,_{,i})^{2}+(-\frac{3l^{2}P_{a}P_{\varphi}}{2a^{2}V^{2}}+a^{3}U_{\varphi})\int d^{3}x\gamma^{\frac{1}{2}}\phi\delta\varphi \nonumber \\
&&-\frac{a}{3l^{2}}\int
d^{3}x\gamma^{\frac{1}{2}}\biggl[\psi^{,i}\psi_{,i}-2\phi^{i}\psi_{i}+K(-3\psi^{2}+6\phi\psi)\biggr]-\frac{aK}{3l^{2}}\int d^{3}x\gamma^{\frac{1}{2}}FF^{,i}\,_{,i}+\frac{1}{2a^{3}}\int d^{3}x\frac{\pi_{\varphi}^{2}}{\gamma^{\frac{1}{2}}} \nonumber \\
&&+\frac{a}{2}\int
d^{3}x\gamma^{\frac{1}{2}}\delta\varphi^{,i}\delta\varphi_{,i}+(-\frac{9l^{2}P_{\varphi}^{2}}{4a^{3}V^{2}}+\frac{a^{3}U_{\varphi\varphi}}{4})\int
d^{3}x\gamma^{\frac{1}{2}}\delta\varphi^{2}
\end{eqnarray}
\end{widetext}

Performing the canonical transformation
\newline
\newline
\begin{eqnarray}
\label{transf2-esc}
\varphi_{0}&=&\tilde{\varphi_{0}}-\frac{1}{V}\int d^{3}x\gamma^{\frac{1}{2}}\tilde{\phi}\tilde{\delta\varphi} \nonumber \\
\pi_{\phi}&=&\tilde{\pi_{\phi}}+\frac{\tilde{P_{\varphi}}}{V}\gamma^{\frac{1}{2}}\tilde{\delta\varphi} \nonumber \\
\pi_{\varphi}&=&\tilde{\pi_{\varphi}}+\frac{\tilde{P_{\varphi}}}{V}\gamma^{\frac{1}{2}}\tilde{\phi}
\end{eqnarray}
generated by
\begin{equation}
\label{gerador2-esc}
\mathcal{F}=\mathcal{I}-\frac{P_{\varphi}}{V}\int
d^{3}x\gamma^{\frac{1}{2}}\tilde{\phi}\delta\varphi,
\end{equation}
where $\mathcal{I}$ represents the identity transformation, the new
$H$ reads
\begin{equation}
\label{h-transf2-escalares-esc} H=N H_{0}+\Lambda_{N}P_{N}+\int
d^{3}x\Lambda_{F}\pi_{F}+\int d^{3}x\Lambda_{\phi}\pi_{\phi},
\end{equation}
where
\begin{widetext}
\begin{eqnarray}
\label{h0-transf2-escalares-esc}
H_{0}&=&-\frac{l^{2}P_{a}^{2}}{4aV}+\frac{P_{\varphi}^{2}}{2a^{3}V}-\frac{KaV}{l^{2}}+\frac{a^{3}VU}{2}+\frac{l^{2}P_{a}}{2a^{2}V}\int
d^{3} \phi\pi_{\psi}+\frac{3l^{2}P_{\varphi}}{2a^{3}V}\int
d^{3}\delta\varphi\pi_{\psi}-\frac{a^{3}}{l^{2}}\int
d^{3}x\gamma^{\frac{1}{2}}(\frac{l^{2}}{2a^{3}\gamma^{\frac{1}{2}}}\pi_{\psi}+\frac{1}{3a}F^{,i}\,_{,i})^{2} \nonumber \\
&&+(-\frac{3l^{2}P_{a}P_{\varphi}}{2a^{2}V^{2}}+\frac{a^{3}U_{\varphi}}{2})\int
d^{3}x\gamma^{\frac{1}{2}}\phi\delta\varphi-\frac{a}{3l^{2}}\int
d^{3}x\gamma^{\frac{1}{2}}\biggl[\psi^{,i}\psi_{,i}-2\phi^{i}\psi_{i}+K(-3\psi^{2}+6\phi\psi)\biggr]-\frac{aK}{3l^{2}}\int
d^{3}x\gamma^{\frac{1}{2}}FF^{,i}\,_{,i} \nonumber \\
&&+\frac{P_{\varphi}}{a^{3}V}\int
d^{3}x\phi\pi_{\varphi}+\frac{1}{2a^{3}}\int
d^{3}x\frac{\pi_{\varphi}^{2}}{\gamma^{\frac{1}{2}}}+\frac{a}{2}\int
d^{3}x\gamma^{\frac{1}{2}}\delta\varphi^{,i}\delta\varphi_{,i}+(-\frac{9l^{2}P_{\varphi}^{2}}{4a^{3}V^{2}}+\frac{a^{3}U_{\varphi\varphi}}{4})\int
d^{3}x\gamma^{\frac{1}{2}}\delta\varphi^{2}
\end{eqnarray}
\end{widetext}

Conservation of the primary constraints
(\ref{vinculos-escalares-esc}) leads to the secondary constraints
\begin{equation}
\label{fi4-esc} H_{0}\approx0
\end{equation}
\begin{equation}
\label{fi5-esc}
\phi_{5}\equiv\frac{1}{3a}\pi_{\psi}+\frac{2a}{9l^{2}}\gamma^{\frac{1}{2}}F^{,i}\,_{,i}+\frac{2aK}{3l^{2}}\gamma^{\frac{1}{2}}F\approx0
\end{equation}
\begin{eqnarray}
\label{fi6-esc} \phi_{6}
&\equiv&-\frac{l^{2}P_{a}}{2a^{2}V}\pi_{\psi}+
(\frac{3l^{2}P_{a}P_{\varphi}}{2a^{2}V^{2}}-
\frac{a^{3}U_{\varphi}}{2})\gamma^{\frac{1}{2}}\delta\varphi
\nonumber \\
&&+\frac{2a}{3l^{2}}\gamma^{\frac{1}{2}}\psi^{,i}\,_{,i}
+\frac{6aK}{3l^{2}}\gamma^{\frac{1}{2}}\psi-\frac{P_{\varphi}}{a^{3}V}\pi_{\varphi}\approx0 \nonumber \\
&&
\end{eqnarray}

Conservation of $H_0$ is identically satisfied. Conservation of
$\phi_{6}$ leads to a term proportional to $H_{0}$ up to second
order terms. Finally, $\phi_{5}$ fixes the Lagrange multiplier
$\Lambda_{F}$:
\begin{equation}
\label{lambdaf-esc} a\Lambda_{F}=\psi-\phi+\frac{l^{2}P_{a}}{aV}F.
\end{equation}
Substituting $\dot{F}=\{F,H\}=\Lambda_F$ into the above
equation, we get for the gauge invariant Bardeen potentials $\Phi$
and $\Psi$
\begin{equation}
\label{fi-psi-esc} \Phi=\Psi
\end{equation}

Calculating the non null Poisson brackets among the constraints
yields
\begin{eqnarray}
\label{parenteses-poisson-esc}
&&\{\phi_{3},\phi_{5}\}=-\frac{2a}{9l^{2}}\gamma^{\frac{1}{2}}\delta^{3}(x-x^{\prime})^{,i}\,_{,i}-
\frac{2aK}{3l^{2}}\gamma^{\frac{1}{2}}\delta^{3}(x-x^{\prime}) \nonumber \\
&&\{\phi_{5},\phi_{6}\}=-\frac{2}{9l^{2}}\gamma^{\frac{1}{2}}\delta^{3}(x-x^{\prime})^{,i}\,_{,i}-
\frac{2K}{3l^{2}}\gamma^{\frac{1}{2}}\delta^{3}(x-x^{\prime}) .\nonumber \\
\end{eqnarray}
The $\phi_{3}$ and $\phi_{5}$ constraints are second class, while
\begin{equation}
\label{fi6-primeiraclasse-esc}
\bar{\phi_{6}}=:\phi_{6}+\frac{1}{a}\phi_{3}
\end{equation}
is a first class constraint. Defining the Dirac brackets, the second
class constraints can be substituted in the hamiltonian.

Making $K=0$, and performing the canonical transformations generated
by
\begin{eqnarray}
\label{gerador3-esc}
\mathcal{F}_{1}&=&a\tilde{P_{a}}+\varphi_{0}\tilde{P_{\varphi}}+\int d^{3}x\biggl[a\pi\delta\varphi+\psi\tilde{\pi_{\psi}}-\frac{2\tilde{P_{\varphi}}}{l^{2}P_{a}}\psi\pi \nonumber \\
&&+\frac{\alpha}{2}\gamma^{\frac{1}{2}}\delta\varphi^{2}\biggr],
\end{eqnarray}
where
\begin{equation}
\label{alfa-esc}
\alpha=\frac{3P_{\varphi}^{2}}{aP_{a}V}+\frac{al^{2}P_{a}}{2V},
\end{equation}
and
\begin{widetext}
\begin{eqnarray}
\label{gerador4-esc-alfa}
\mathcal{F}_{2}&=&a\tilde{P_{a}}+\varphi_{0}\tilde{P_{\varphi}}+\int
d^{3}x\biggl\{\psi\tilde{\pi_{\psi}}+v\tilde{\pi}+(\frac{2\tilde{P_{\varphi}}}{aV}-\frac{a^{4}VU_{\varphi}}{l^{2}\tilde{P_{a}}}-\frac{6\tilde{P_{\varphi}}^{3}}{l^{2}a^{3}\tilde{P_{a}}^{2}V})\gamma^{\frac{1}{2}}v\psi+(\frac{2\tilde{P_{\varphi}}^{2}}{l^{2}a\tilde{P_{a}}V}-\frac{a^{4}VU_{\varphi}\tilde{P_{\varphi}}}{l^{4}\tilde{P_{a}}^{2}}-\frac{6\tilde{P_{\varphi}}^{4}}{l^{4}a^{3}\tilde{P_{a}}^{3}V})\gamma^{\frac{1}{2}}\psi^{2} \nonumber \\
&&+\frac{2a^{3}V}{3l^{4}\tilde{P_{a}}}\gamma^{\frac{1}{2}}\psi\psi^{,i}\,_{,i}\biggr\},
\end{eqnarray}

and making a redefinition of $N$, we finally obtain
\begin{equation}
\tilde{\phi_{6}}=\pi_{\psi},
\end{equation}
and
\begin{eqnarray}
H&=&NH_{0}+\int
d^{3}x\biggl(-\frac{l^{2}P_{a}}{2a^{2}V}\phi+\frac{3P_{\varphi}^{2}}{a^{4}P_{a}V}\psi+\frac{3l^{2}P_{\varphi}}{2a^{4}V}v\biggr)\tilde{\phi_{6}}+\Lambda_{N}P_{N}+\int
d^{3}x \Lambda_{\phi}\pi_{\phi},
\end{eqnarray}
where
\begin{eqnarray}
\label{h0-simplif-esc}
H_{0}&=&-\frac{l^{2}P_{a}^{2}}{4aV}+\frac{P_{\varphi}^{2}}{2a^{3}V}+\frac{a^{3}VU}{2}+\frac{1}{2a}\int d^{3}x \frac{\pi^{2}}{\gamma^{\frac{1}{2}}}+\frac{1}{2a}\int d^{3}x\gamma^{\frac{1}{2}}v^{,i}v_{,i} \nonumber \\
&&+\biggl(\frac{15l^{2}P_{\varphi}^{2}}{4a^{5}V^{2}}+\frac{aU_{\varphi\varphi}}{4}-\frac{3l^{2}Ua}{8}+\frac{9UP_{\varphi}^{2}}{4aP_{a}^{2}}-\frac{l^{4}P_{a}^{2}}{16a^{3}V^{2}}-\frac{27P_{\varphi}^{4}}{4a^{7}V^{2}P_{a}^{2}}-\frac{3P_{\varphi}U_{\varphi}}{P_{a}}\biggr)\int
d^{3}x\gamma^{\frac{1}{2}}v^{2}.
\end{eqnarray}
Using the background classical equations one can show that the
coefficient of $v^{2}$ can be written as $z''/z$ as in \cite{MFB}.
Without their use, this is the simplest form the hamiltonian of
scalar perturbations can have in scalar field models.

\section{The explicit canonical transformations}

The explicit canonical transformations obtained from the generators
$\mathcal{F}_1$, $\mathcal{F}_2$ and $\mathcal{F}_3$ of section III are, respectively

\begin{eqnarray}
\label{transf2-27hidro}
a&=&\tilde{a}\biggl[1+\frac{2\sqrt{(\lambda+1)P_{T}}\sqrt{V}}{l^{2}\tilde{P}_{a}^{2}\sqrt{\lambda}}\tilde{a}^{\frac{1}{2}(1-3\lambda)}\int
d^{3}x\tilde{\psi}\pi+\frac{1}{2\tilde{a}}\frac{\partial\alpha}{\partial
\tilde{P}_{a}}\int
d^{3}x\gamma^{\frac{1}{2}}(\sqrt{6}l\tilde{a}^{\frac{1}{2}(1-3\lambda)}v-\frac{2\sqrt{6}\sqrt{(\lambda+1)P_{T}}\sqrt{V}}{l\tilde{P}_{a}\sqrt{\lambda}}\tilde{a}^{2-3\lambda}\tilde{\psi})^{2}\biggr]
\nonumber \\
P_{a}&=&\tilde{P}_{a}-\frac{(1-3\lambda)}{2\sqrt{6}l}\tilde{a}^{-\frac{3}{2}(1-\lambda)}\int
d^{3}x(\sqrt{6}l\tilde{a}^{\frac{1}{2}(1-3\lambda)}v-\frac{2\sqrt{6}\sqrt{(\lambda+1)P_{T}}\sqrt{V}}{l\tilde{P}_{a}\sqrt{\lambda}}\tilde{a}^{2-3\lambda}\tilde{\psi)}\pi \nonumber \\
&&+\frac{3(1-\lambda)\sqrt{(\lambda+1)P_{T}}\sqrt{V}}{l^{2}\tilde{P}_{a}\sqrt{\lambda}}\tilde{a}^{\frac{1}{2}(1-3\lambda)}\int
d^{3}x\tilde{\psi}\pi-\frac{1}{2}\frac{\partial\alpha}{\partial a}\int d^{3}x\gamma^{\frac{1}{2}}(\sqrt{6}l\tilde{a}^{\frac{1}{2}(1-3\lambda)}v-\frac{2\sqrt{6}\sqrt{(\lambda+1)P_{T}}\sqrt{V}}{l\tilde{P}_{a}\sqrt{\lambda}}\tilde{a}^{2-3\lambda}\tilde{\psi})^{2} \nonumber \\
\pi_{\varphi}\,_{(c)}&=&\frac{\tilde{a}^{-\frac{1}{2}}(1-3\lambda)}{\sqrt{6}l}\pi-\alpha\sqrt{6}l\tilde{a}^{\frac{1}{2}(1-3\lambda)}\gamma^{\frac{1}{2}}v+\frac{2\alpha\sqrt{6}\sqrt{(\lambda+1)P_{T}}\sqrt{V}}{l\tilde{P}_{a}\sqrt{\lambda}}\tilde{a}^{2-3\lambda}\gamma^{\frac{1}{2}}\tilde{\psi} \nonumber \\
\varphi_{(c)}&=&\sqrt{6}l\tilde{a}^{\frac{1}{2}(1-3\lambda)}v-\frac{2\sqrt{6}\sqrt{(\lambda+1)P_{T}}\sqrt{V}}{l\tilde{P}_{a}\sqrt{\lambda}}\tilde{a}^{2-3\lambda}\tilde{\psi} \nonumber \\
\pi_{\psi}\,_{(c)}&=&\tilde{\pi}_{\psi}+\frac{2\sqrt{(\lambda+1)P_{T}}\sqrt{V}}{l^{2}\tilde{P}_{a}\sqrt{\lambda}}\tilde{a}^{\frac{3}{2}(1-\lambda)}\pi,
\nonumber \\
&&
\end{eqnarray}
\newline
\newline
\newline
\begin{eqnarray}
\label{transf3-27hidro}
a&=&\tilde{a}\biggl\{1+\frac{12\sqrt{V}[(\lambda+1)P_{T}]^{\frac{3}{2}}}{l^{2}\tilde{P}_{a}^{3}\sqrt{\lambda}}\tilde{a}^{\frac{1}{2}(91-9\lambda)}\int
d^{3}x\gamma^{\frac{1}{2}}v\tilde{\psi}+\biggl[-\frac{18V[(\lambda+1)P_{T}]^{2}}{l^{4}\tilde{P}_{a}^{4}\lambda}\tilde{a}^{2-6\lambda}+\frac{(1-3\lambda)(\lambda1)P_{T}}{2l^{2}\tilde{P}_{a}^{2}\lambda}\tilde{a}^{1-3\lambda}\biggr]\int
d^{3}x\gamma^{\frac{1}{2}}\psi^{2} \nonumber \\
&&+\frac{2\tilde{a}^{2}V}{3l^{4}\tilde{P}_{a}^{2}}\int
d^{3}x\gamma^{\frac{1}{2}}\tilde{\psi}\tilde{\psi}^{,i}\,_{,i}\biggr\} \nonumber \\
P_{a}&=&\tilde{P}_{a}+\biggl[\frac{9(1-3\lambda)\sqrt{V}[(\lambda+1)P_{T}]^{\frac{3}{2}}}{l^{2}\tilde{P}_{a}^{2}\sqrt{\lambda}}\tilde{a}^{\frac{1}{2}(1-9\lambda)}-\frac{(1-9\lambda^{2})\sqrt{(\lambda+1)P_{T}}}{4\sqrt{\lambda}\sqrt{V}}\tilde{a}^{-\frac{1}{2}(1+3\lambda)}\biggr]\int d^{3}x\gamma^{\frac{1}{2}}\tilde{v}\tilde{\psi} \nonumber \\
&&+\biggl[-\frac{18V(1-2\lambda)[(\lambda+1)P_{T}]^{2}}{l^{4}\tilde{P}_{a}^{3}\lambda}\tilde{a}^{2-6\lambda}+\frac{(2+3\lambda-9\lambda^{2})(\lambda+1)P_{T}}{2l^{2}\tilde{P}_{a}\lambda}\tilde{a}^{1-3\lambda}\biggr]\int
d^{3}x\gamma^{\frac{1}{2}}\tilde{\psi}^{2}+\frac{2\tilde{a}^{2}V}{l^{4}\tilde{P}_{a}}\int d^{3}x\gamma^{\frac{1}{2}}\tilde{\psi}\tilde{\psi}^{,i}\,_{,i} \nonumber \\
\pi&=&\tilde{\pi}+\biggl[\frac{6\sqrt{V}[(\lambda+1)P_{T}]^{\frac{3}{2}}}{l^{2}\tilde{P}_{a}^{2}\sqrt{\lambda}}\tilde{a}^{\frac{3}{2}(1-3\lambda)}-\frac{(1+3\lambda)\sqrt{(\lambda+1)P_{T}}}{2\sqrt{\lambda}\sqrt{V}}\tilde{a}^{\frac{1}{2}(1-3\lambda)}\biggr]\gamma^{\frac{1}{2}}\tilde{\psi} \nonumber \\
\pi_{\psi}&=&\tilde{\pi_{\psi}}+\biggl[\frac{6\sqrt{V}[(\lambda+1)P_{T}]^{\frac{3}{2}}}{l^{2}\tilde{P}_{a}^{2}\sqrt{\lambda}}\tilde{a}^{\frac{3}{2}(1-3\lambda)}-\frac{(1+3\lambda)\sqrt{(\lambda+1)P_{T}}}{2\sqrt{\lambda}\sqrt{V}}\tilde{a}^{\frac{1}{2}(1-3\lambda)}\biggr]\gamma^{\frac{1}{2}}\tilde{v} \nonumber \\
&&+\biggl[-\frac{12V[(\lambda+1)P_{T}]^{2}}{l^{4}\tilde{P}_{a}^{3}\lambda}\tilde{a}^3{1-2\lambda}+\frac{(1+3\lambda)(\lambda+1)P_{T}}{l^{2}\tilde{P}_{a}\lambda}\tilde{a}^{2-3\lambda}\biggr]\gamma^{\frac{1}{2}}\tilde{\psi}+\frac{4\tilde{a}^{3}V}{3l^{4}\tilde{P}_{a}}\gamma^{\frac{1}{2}}\tilde{\psi}^{i}\,_{i},
\end{eqnarray}
\end{widetext}

\begin{eqnarray}
\label{transf4-hidro}
a&=&\tilde{a}+\frac{l^{2}}{4\tilde{a}V}\int d^{3}x\gamma^{\frac{1}{2}}\tilde{v}^{2} \nonumber \\
P_{a}&=&\tilde{P}_{a}-\frac{1}{\tilde{a}^{2}}\int
d^{3}xv\pi+\frac{l^{2}\tilde{P}_{a}}{4V}\int
d^{3}\gamma^{\frac{1}{2}}\tilde{v}^{2} \nonumber \\
\pi&=&\frac{1}{a}\tilde{\pi}-\frac{l^{2}\tilde{P}_{a}}{2V}\gamma^{\frac{1}{2}}\tilde{v}
\nonumber \\ v&=&a\tilde{v}.
\end{eqnarray}

The intermediary hamiltonian between $\mathcal{F}_2$ and
$\mathcal{F}_3$ reads,
\begin{widetext}
\begin{eqnarray}
\label{h0-transf3-hidro}
H_{0}&=&-\frac{l^{2}P_{a}^{2}}{4aV}+\frac{P_{T}}{a^{3\lambda}}+\frac{1}{2a}\int
d^{3}x\frac{\pi^{2}}{\gamma^{\frac{1}{2}}}+\frac{\lambda}{2a}\int
d^{3}x\gamma^{\frac{1}{2}}v^{,i}v_{,i}-\frac{(1-3\lambda)l^{2}P_{T}}{4V}a^{-(2+3\lambda)}v^{2} \nonumber \\
&&+\biggl\{-\frac{\sqrt{\lambda}}{2}\frac{3l^{2}\sqrt{(\lambda+1)P_{T}}}{\sqrt{V}}a^{-\frac{1}{2}(5+3\lambda)}v+\frac{3(\lambda+1)P_{T}}{P_{a}}a^{-(1+3\lambda)}\psi\biggr\}\pi_{\psi}+H_{0}^{(0)}\int
d^{3}x\biggl\{-\frac{9(\lambda+1)P_{T}}{l^{2}P_{a}^{2}}a^{1-3\lambda}\gamma^{\frac{1}{2}}\psi^{2} \nonumber \\
&&+\frac{6\sqrt{V}\sqrt{(\lambda+1)P_{T}}}{l^{2}P_{a}^{2}\sqrt{\lambda}}a^{\frac{1}{2}(1-3\lambda)}\psi\pi+\biggl[\frac{18[(\lambda+1)P_{T}]^{\frac{3}{2}}\sqrt{V}}{l^{2}P_{a}^{3}\sqrt{\lambda}}a^{\frac{1}{2}(1-9\lambda)}-\frac{3(1-3\lambda)\sqrt{(\lambda+1)P_{T}}}{2\sqrt{V}\sqrt{\lambda}P_{a}}a^{-\frac{1}{2}(1+3\lambda)}\biggr]\gamma^{\frac{1}{2}}v\psi \nonumber \\
&&+\frac{2a^{2}V}{l^{4}P_{a}^{2}}\psi\psi^{,i}\,_{,i}+\biggl[-\frac{9(1+2\lambda)[(\lambda+1)P_{T}]}{2P_{a}^{2}}a^{-(1+3\lambda)}+\frac{(2-9\lambda+9\lambda^{2})l^{2}}{8Va^{2}}\biggr]\gamma^{\frac{1}{2}}v^{2}\biggr\}.
\end{eqnarray}
\end{widetext}

\end{document}